# DeepMB: Deep neural network for real-time optoacoustic image reconstruction with adjustable speed of sound

Christoph Dehner[1,2,*], Guillaume Zahnd[1,3,*], Vasilis Ntziachristos[1,2,4,+], Dominik Jüstel[1,2,5,+]

[1]*Institute of Biological and Medical Imaging, Helmholtz Zentrum München, Neuherberg, Germany*
[2]*Chair of Biological Imaging at the Central Institute for Translational Cancer Research (TranslaTUM), School of Medicine, Technical University of Munich, Germany*
[3]*iThera Medical GmbH, Munich, Germany*
[4]*Munich Institute of Robotics and Machine Intelligence, Technical University of Munich, Germany*
[5]*Institute of Computational Biology, Helmholtz Zentrum München, Neuherberg, Germany*

*\*These authors contributed equally to this work.*
*+These authors share the corresponding authorship.*

**Abstract:** Multispectral optoacoustic tomography (MSOT) is a high-resolution functional imaging modality that can non-invasively access a broad range of pathophysiological phenomena by quantifying the contrast of endogenous chromophores in tissue. Real-time imaging is imperative to translate MSOT into clinical imaging, visualize dynamic pathophysiological changes associated with disease progression, and enable in situ diagnoses. Model-based reconstruction affords state-of-the-art optoacoustic images; however, the advanced image quality provided by model-based reconstruction remains inaccessible during real-time imaging because the algorithm is iterative and computationally demanding. Deep learning affords faster reconstruction of optoacoustic images, but the lack of experimental ground truth training data can lead to reduced image quality for in vivo data. In this work, we achieve accurate optoacoustic image reconstruction for arbitrary (experimental) input data in 31 ms per image by expressing model-based reconstruction with a deep neural network. The proposed deep learning framework, termed DeepMB, facilitates accurate generalization from synthesized training data to experimental test data through training on optoacoustic signals synthesized from real-world images and ground truth optoacoustic images generated by model-based reconstruction. The framework affords in-focus images for a broad range of anatomical locations with various acoustic properties because it supports dynamic adjustment of the reconstruction speed of sound during imaging. Furthermore, DeepMB is compatible with the data rates and image sizes of modern multispectral optoacoustic tomography scanners, thus enabling straightforward adoption into clinical routine. We evaluate DeepMB both qualitatively and quantitatively on a diverse dataset of in vivo images and demonstrate that the framework reconstructs images approximately 1000 times faster than the iterative model-based reference method while affording near-identical image qualities. Accurate and real-time image reconstructions with DeepMB can enable full access to the high-resolution and multispectral contrast of handheld optoacoustic tomography in deep tissue, thus facilitating advanced dynamic imaging applications.

## 1. Introduction

Multispectral optoacoustic tomography (MSOT) is an emerging functional imaging modality that uniquely enables non-invasive detection of optical contrast at high spatial resolution and centimeter-scale penetration depth in living tissue[1-6]. Accessing the multispectral contrast of endogenous chromophores, MSOT can quantify a broad range of pathophysiological surrogate biomarkers such as tissue fibrosis, inflammation, vascularization, and oxygenation, and provide unmatched clinical information for multifarious diseases such as breast cancer[2, 6], Duchenne muscular dystrophy[7], or inflammatory bowel disease[3].

In order to fully translate and integrate MSOT into clinical imaging, real-time application is imperative[8-10]. Handheld MSOT imaging requires — similar to ultrasound imaging — live image feedback at sufficiently high frame-rates (at least 24 fps for full-video rendering) to avoid hindering visio-tactile coordination, identify and localize relevant tissue structures using anatomical landmarks in their surroundings, and find the optimal transducer pose for the target region. Furthermore, real-time optoacoustic imaging is necessary to visualize dynamic pathophysiological changes associated with disease progression and enable in situ guidance and diagnosis during intra-operative and endoscopy imaging[11, 12]. In practice, real-time reconstruction of optoacoustic images (i.e., recovery of the initial pressure distribution in the imaged tissue) is generally conducted via the backprojection algorithm[13].

However, the backprojection formula is based on over-simplified modelling assumptions of the imaging process and cannot compensate for the ill-posedness of the underlying inverse problem arising from limited-angle acquisition, measurement noise, and finite transducer bandwidth. Consequently, backprojection images systematically suffer from low spatial resolution and contrast, as well as negative pixel values that invalidate a physical interpretation of the image as initial pressure distribution. In contrast, iterative model-based reconstruction[14, 15] can provide accurate optoacoustic images with state-of-the-art quality by incorporating a physical model of the imaging device into the reconstruction process, constraining the reconstructed image to be non-negative, and introducing regularization to mitigate the ill-posedness of the inversion problem. However, model-based reconstruction is computationally demanding because of the iterative and thus sequential nature of the algorithm, which is prohibitive for real-time imaging. Real-time model-based reconstruction has been demonstrated for a pre-clinical MSOT system by computing the reconstruction on a graphics processing unit (GPU)[16]. However, a similar acceleration is infeasible for state-of-the-art model-based reconstruction of data from modern clinical systems because these reconstructions are substantially more computationally demanding (larger images, more complex regularization functionals, inclusion of the total impulse response of the system in the model, necessity of a higher number of iterations until convergence[14, 15]). Therefore, the full imaging potential of MSOT is only available offline after considerable computational time and currently remains inaccessible for clinical applications that require live image feedback.

Recently, deep neural networks have been successfully applied to various inverse problems in imaging, utilizing their ability to capture suitable inverse transforms in a data-driven way and efficiently apply these transforms to new data[17-23]. Real-time image reconstruction with deep learning has been achieved using deep loop unfolding and direct inference. Deep loop unfolding involves interpreting the iterations of a variational reconstruction algorithm as the layers of a convolutional neural network, and training the resulting network end-to-end in a supervised fashion[24-28]. This methodology has been shown to facilitate accurate and efficient image reconstruction for various medical imaging modalities such as magnetic resonance imaging, computed tomography, or intensity diffraction tomography. However, deep loop unfolding is unsuited for real-time optoacoustic image reconstruction because it requires repeated evaluations of the involved optoacoustic forward model (at least one forward and one adjoint model evaluation per data consistency block, see e.g. equation 11 in MoDL framework[24]), which is too computationally expensive to enable real-time processing (e.g., with the imaging setup from this paper, a single evaluation of the forward or adjoint model already takes more than 50 ms on a NVIDIA GeForce RTX3090 GPU). Conversely, deep-learning-based image reconstruction via direct inference can support real-time optoacoustic imaging because the approach does not require to evaluate the optoacoustic forward model during image reconstruction. Over the past few years, several direct inference methods have been introduced to either directly infer high-quality images from recorded signals[29-35] or accelerate the minimization operation from iterative model-based reconstruction[36].

A key challenge in applying deep learning for optoacoustic image reconstruction is the generation of appropriate training data, i.e., input sinograms and corresponding optoacoustic initial pressure reference images. In general, network training must rely on synthetized data because ground truth information about the initial pressure distribution in biological tissue is not available experimentally. Data synthesis involves hand-crafting reference distributions of the initial pressure and simulating the corresponding input sinograms using a physical forward model of the imaging process. However, such synthesized sinograms and reference images only partially represent the true properties of experimental data, hence their use as input-target pairs for network training can lead to reductions in reconstruction accuracy for in vivo data.

In this work, we show that learning a well-posed reconstruction operator facilitates accurate generalization from synthesized training data to experimental test data. We achieve real-time optoacoustic image reconstruction for arbitrary (experimental) input data by expressing model-based reconstruction with a deep neural network. The proposed deep learning framework, DeepMB, learns an accurate and universally applicable model-based optoacoustic reconstruction operator through training on optoacoustic signals synthesized from real-world images, while using as ground truth the optoacoustic images generated by model-based reconstruction of the corresponding signals. DeepMB affords image quality nearly-indistinguishable from state-of-the-art iterative model-based reconstructions at speeds enabling live imaging (32 fps, or 31 ms/image, versus 30-60 s/image for iterative model-based reconstruction). Furthermore, DeepMB is directly compatible with state-of-the-art clinical MSOT scanners because it supports high throughput data acquisition (sampling rate: 40 MHz; number of transducers: 256) and large image sizes (416×416 pixels). DeepMB also supports dynamic adjustments of the SoS parameter during imaging, which enables the reconstruction of in-focus images for arbitrary tissue types. We demonstrate the performance of DeepMB both quantitatively and qualitatively on a diverse dataset of in vivo images (4814 images, 6 participants, 25-29 scanned locations per participant). With DeepMB, clinical MSOT could provide high quality feedback during live imaging and thus facilitate advanced dynamic imaging applications.

## 2. Results

To validate the capability of DeepMB to reconstruct images in real-time and with adjustable SoS, the framework was applied to a modern handheld optoacoustic scanner (MSOT Acuity Echo, iThera Medical GmbH, Munich, Germany) with SoS values ranging from 1475 m/s to 1525 m/s in steps of 5 m/s.

**DeepMB pipeline**

Fig. 1 illustrates the overall training and evaluation pipeline. DeepMB was trained, similarly to the AUTOMAP framework[18], using input sinograms synthesized from general-feature images to facilitate the learning of an unbiased and universally applicable reconstruction operator. These sinograms were generated by employing a diverse collection of publicly available real-world images[37] as initial pressure distributions and simulating thereof the signals recorded by the acoustic transducers with an accurate physical model of the considered scanner[14] (Fig. 1a and section "Methods / Synthesis of sinograms for training and validation"). The SoS values for the forward simulations were drawn uniformly at random from the considered range for each image. Ground-truth images for the synthesized sinograms were computed via model-based reconstruction (Fig. 1c). Fig. 1d shows the deep neural network architecture of DeepMB, which inputs a sinogram (either synthetic or in vivo) and an SoS value and outputs the final reconstructed image. The underlying design is based on the U-Net architecture[38] augmented with two extensions that promote the network to learn and express the effects of the different input SoS values onto the reconstructed images: (1) all signals were mapped from the input sinogram to the image domain with a linear delay operator based on the given input SoS value (no trainable weights), and (2) the input SoS value (one-hot encoded and concatenated as additional channels) was passed to the trainable convolutional layers of the network. A detailed description of the network training is given in the section "Methods / Network training". After training, the applicability of DeepMB to clinical data was tested with a diverse dataset of in vivo sinograms acquired by scanning six participants at up to eight anatomical locations each (Fig. 1b). The corresponding ground-truth images of the acquired in vivo test sinograms were obtained analogously to the training data via model-based reconstruction. The inference time of DeepMB was 31 ms per sample on a modern GPU (NVIDIA GeForce RTX 3090).

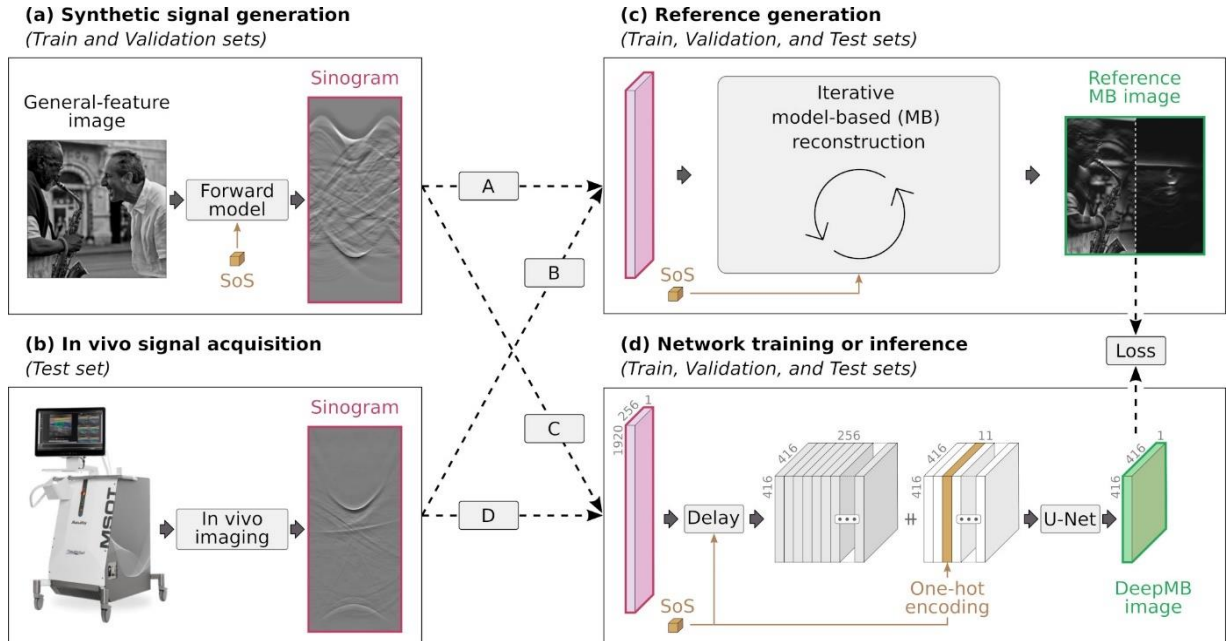

Figure 1: DeepMB pipeline. (a) Real-world images, obtained from a publicly available dataset, are used to generate synthetic sinograms by applying an accurate physical forward model of the scanner. SoS, speed of sound. (b) In vivo sinograms are acquired from diverse anatomical locations in six participants. (c) Optoacoustic images are reconstructed via iterative model-based reconstruction for the purpose of generating reference images for either the synthetic dataset (A) or the in vivo dataset (B). (d) Network training is conducted by using the synthetic data as sets for training (n=8000) and validation (n=2000) (C), while the in vivo data constitutes the test set (n=4814) (D). A domain transformation is first applied to the input sinograms via a delay operation to map the time sample values into the image space. The SoS is then one-hot encoded and concatenated as additional channels (represented by the symbol "⧺"). A U-Net convolutional neural network is subsequently applied to the channel stack to regress the final image. The loss is calculated between the network output and the corresponding reference image (see section "Methods / Network training" for further details about the network training).

## Qualitative evaluation

DeepMB successfully reconstructed high-quality optoacoustic images. To qualitatively evaluate DeepMB, all DeepMB images from the in vivo dataset (Fig. 1b) were thoroughly compared to their corresponding model-based reference images (Fig. 1c). Fig. 2 shows four reconstructed images, corresponding to scans of the carotid artery, biceps, breast, and abdomen. DeepMB reconstructions (Fig. 2a–d) are systematically nearly-indistinguishable from the model-based references (Fig. 2e–h), with no noticeable failures, outliers, or artifacts for any of the participants, anatomies, probe orientations, SoS values, or laser wavelengths. The similarity between the DeepMB and model-based images is also confirmed by their negligible pixel-wise absolute differences (Fig. 2i–l). The zoomed region D in Fig. 2j depicts one of the largest observed discrepancies between the DeepMB and model-based reconstructions, which manifests as minor blurring, showing that the DeepMB image is only marginally affected by these errors. In comparison, backprojection images (Fig. 2m–p) exhibit notable differences from the reference model-based images and suffer from reduced spatial resolution and physically-nonsensical negative initial pressure values. Finally, to facilitate relating the reconstructed optoacoustic images to the scanned anatomies, Fig. 2q–t depict sketches of the anatomical context for all scans, while Fig. 2u–x depict the interleaved-acquired ultrasound images overlayed with the temporally-corresponding DeepMB reconstructions. Extended Data Figs. 1-2 complement the qualitative comparison from Figure 2: Extended Data Fig. 1 shows that the image quality of DeepMB is also superior to the backprojetion algorithm with negative values set to zero after the reconstruction, as well as to the delay-multiply-and-sum with coherence factor algorithm[39, 40]; and Extended Data Fig. 2 shows that DeepMB images are nearly-indistinguishable from model-based references in the case of both very high and very low data residual norms.

Extended Data Videos 1-2 further illustrate the real-time optoacoustic imaging capabilities of DeepMB. Extended Data Video 1 shows a carotid artery continuously imaged in the transversal view at 800 nm, which demonstrates that DeepMB can be used to visualize motion at 25 Hz with state-of-the-art image quality. Extended Data Video 2 shows the optoacoustic image of a biceps in the transversal view at 800 nm while the SoS is gradually adjusted via a series of DeepMB reconstructions, which illustrates the importance of on-the-fly SoS tuning for optimal image quality.

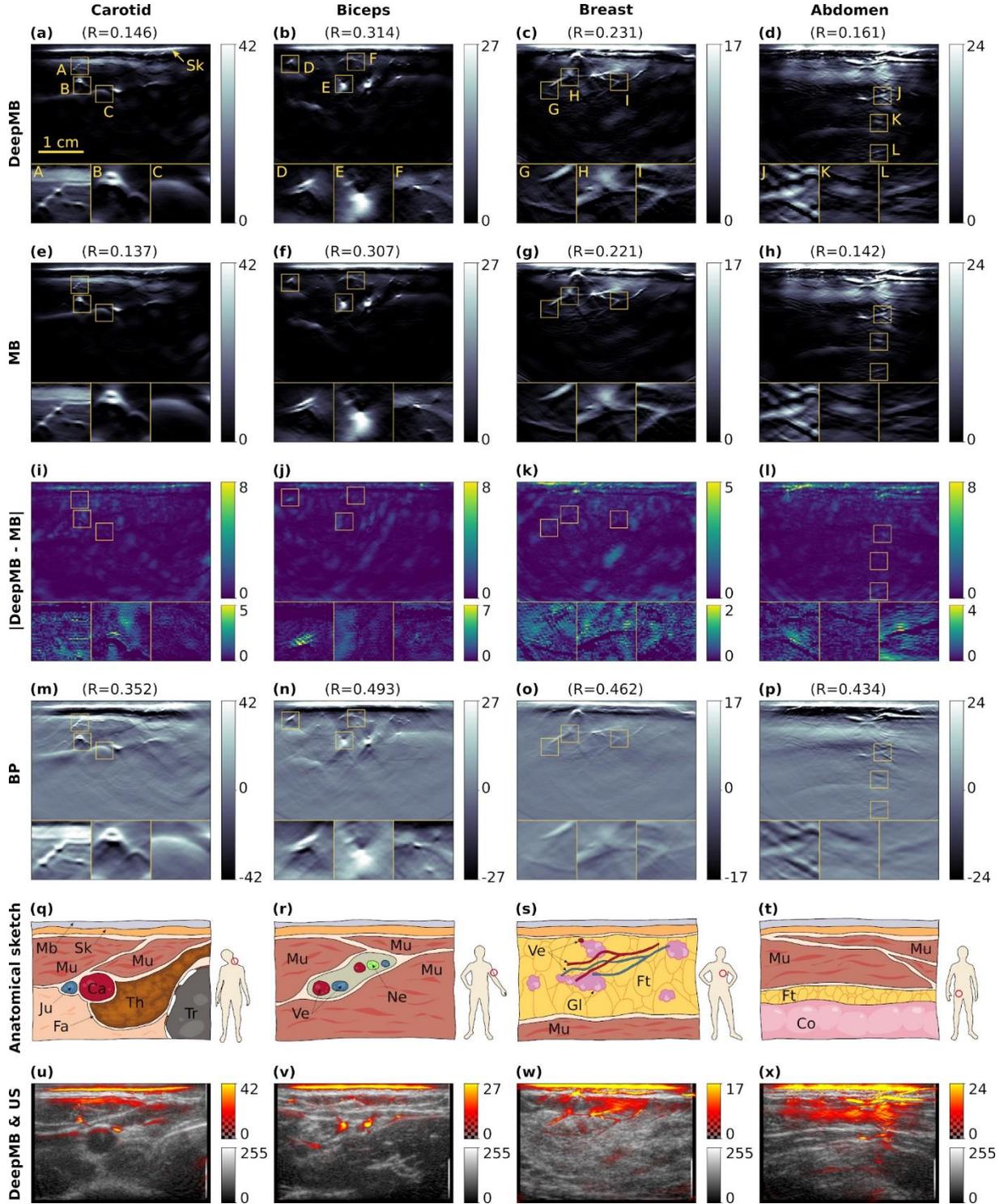

Figure 2: Examples from the in vivo test dataset for different anatomical locations (carotid artery: a,e,i,m,q,u; biceps: b,f,j,n,r,v; breast: c,g,k,o,s,w; abdomen: d,h,l,p,t,x). The first four rows show deep model-based (DeepMB) reconstructions, model-based (MB) reconstructions, the pixel-wise absolute difference between DeepMB and MB reconstructions, and backprojection (BP) reconstructions, respectively. The data residual norm (R) is indicated between round brackets above each image. The last two rows display sketches of the anatomical context of the scans and the interleaved-acquired ultrasound (US) images overlayed with DeepMB reconstructions, respectively.

All optoacoustic images and difference maps show the reconstructed initial pressure in arbitrary units and were slightly cropped to a field of view of 4.16×2.80 cm$^2$ to disregard the area occupied by the probe couplant above the skin line. Each enlarged region is 0.41×0.41 cm$^2$ and displays various anatomical details. All displayed scans were acquired at 800 nm. Mb: probe membrane, Sk: skin, Mu: muscle, Fa: fascia, Ca: common carotid artery, Ju: jugular vein, Th: thyroid, Tr: trachea, Ve: blood vessel, Ne: nerve, Ft: fat, Gl: glandular tissues, Co: colon.

## Quantitative evaluation

The ability of DeepMB to reconstruct images with equivalent fidelity to those afforded by model-based reconstruction was then confirmed by quantitative comparison. To quantify the image fidelity of DeepMB reconstructions, the data residual norm was calculated for all in vivo test images (see section "Methods / Data residual norm" for the precise definition). The data residual norm measures the fidelity of a reconstructed image by computing the mismatch between the image and the corresponding recorded acoustic signals with regard to the accurate physical forward model of the used scanner, and is mathematically-proven minimal for model-based reconstruction[41]. The data residual norm was also calculated for all model-based and backprojection reconstructions, for comparison purposes.

First, data residual norms were calculated with in-focus images (that is, reconstructed with optimal SoS values) to evaluate the fidelity of DeepMB images with the best possible quality (Fig. 3a). Data residual norms of DeepMB images (green, mean±std = 0.156±0.088) are almost as low as the data residual norms of model-based images (blue, mean±std = 0.139±0.095). The close agreement between data residual norms of DeepMB and model-based images confirms that both reconstruction approaches afford equivalent image qualities. In contrast, the data residual norms of backprojection images are markedly higher (gray, mean±std = 0.369±0.098), which reaffirms the shortcomings of backprojection to accurately model the imaging process, and explains the lower image quality observed in Fig. 2d,h,l,p. Table 1 summarizes the data residual norms of all reconstruction approaches evaluated in this paper. Extended Data Table 1 complements the quantitative comparison from Table 1 and confirms that the data residual norms of DeepMB images are almost as low as data residual norms of model-based images even when aggregated separately based on anatomical regions, participants, Fitzpatrick scale, body type, wavelength, and SoS values.

Second, data residual norms were calculated for out-of-focus images (that is, reconstructed with sub-optimal SoS values) to evaluate the fidelity of DeepMB images during imaging applications with a priori unknown SoS (Fig. 3b, also see Table 1). Data residual norms of DeepMB images remain close to those of model-based images for all considered levels of mismatch between the optimal and the employed SoS, thus confirming that DeepMB and model-based images are similarly trustworthy independent of the selected SoS. Note that the two rightmost distributions of data residual norms in Fig. 3b get narrower and include less extreme data residual norm values because they contain fewer data points.

In addition to the quantitative evaluation with data residual norms, the deviation of DeepMB and backprojection images from reference model-based reconstructions were also quantified by computing the mean absolute error (MAE), the relative mean absolute error (MAE$_{rel}$), the mean squared error (MSE), the relative mean squared error (MSE$_{rel}$), and the structural similarity index (SSIM). The obtained metrics for the in vivo test scans are reported in Table 1 and confirm that DeepMB images are very similar to model-based images, whereas backprojection images notably differ from the model-based references.

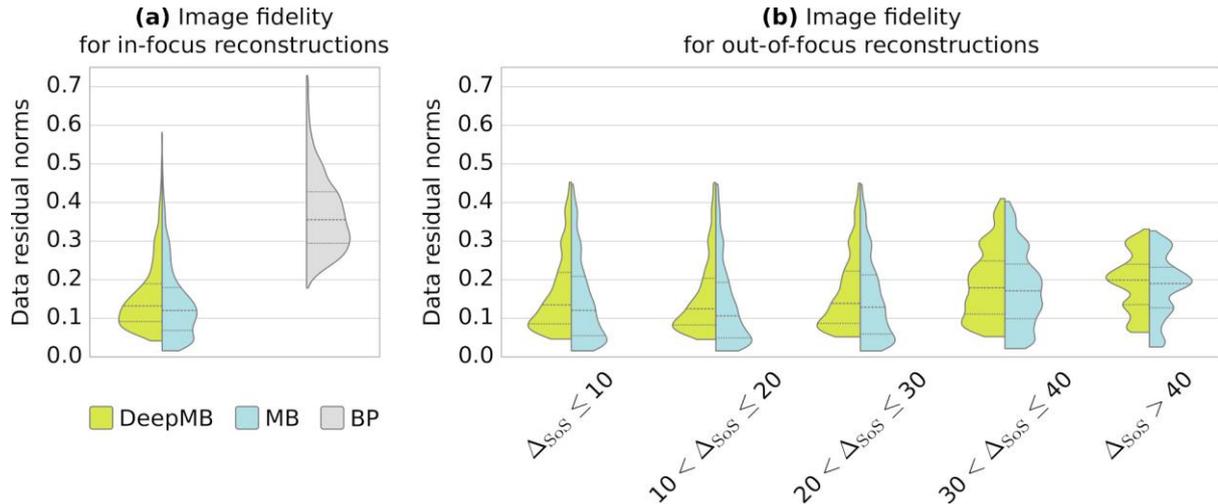

Figure 3: Data residual norms of optoacoustic images from deep model-based (DeepMB), model-based (MB), and backprojection (BP) reconstruction. (a) Data residual norms of in-focus images reconstructed with optimal speed of sound (SoS) values, on all 4814 samples from the in vivo test set. (b) Data residual norms of out-of-focus images reconstructed with sub-optimal SoS values, on a subset of 638 samples. The five sub-panels depict the effect of SoS mismatch via gradual increase of the offset ΔSoS in steps of 10 m/s. The inner bars indicate the 25th, 50th, and 75th percentiles.

**Multispectral evaluation**

The previously described experiments validate, with in vivo scans from the range 700–980 nm, that the single-wavelength image quality of DeepMB is nearly-identical to model-based reconstruction and clearly superior to backprojection reconstruction. Additional experiments were then conducted to show that the multispectral image contrast of DeepMB is comparable to model-based reconstruction and superior to backprojection reconstruction.

To evaluate the multispectral image quality of DeepMB, model-based, and backprojection reconstruction, all in-vivo scans from the test dataset were grouped into multispectral stacks of 29 images (respectively one scan from the range 700–980 nm in steps of 10 nm, see section "Methods / Acquisition of in vivo test sinograms") and linearly unmixed into oxyhemoglobin, deoxyhemoglobin, fat, and water components[42]. Fig. 4 visualizes the unmixed components from DeepMB, model-based, and backproejction images for a representative breast scan, showing in Fig. 4a-c the unmixed components for fat and water, in Fig. 4d-f the unmixed components for oxyhemoglobin and deoxyhemoglobin, in Fig. 4g the reference absorption spectra of the four chromophores used during unmixing, and in Fig. 4h a schematic sketch of the anatomical context for the depicted scan. The unmixed DeepMB images (Fig.4a,d) are systematically nearly-indistinguishable from the model-based references (Fig.4a,d). Conversely, the unmixed backprojection images (Fig. 4c,f) exhibit considerably lower multispectral contrast (see for example the magnifications A-C in Fig. 4c) and miss important image structures (see for example the fine vascularity in magnification B of Fig. 4f). Extended Data Figs. 3-5 visualize the unmixing results of three further in vivo scans and also display unmixed images from the delay-multiply-and-sum with coherence factor algorithm. Finally, the ability of DeepMB to obtain clearly superior multispectral images as backprojection and delay-multiply-and-sum with coherence factor was confirmed quantitatively by computing the structural similarity index, mean squared error, and mean absolute error for all unmixed images against the reference unmixed model-based images (see Table 2).

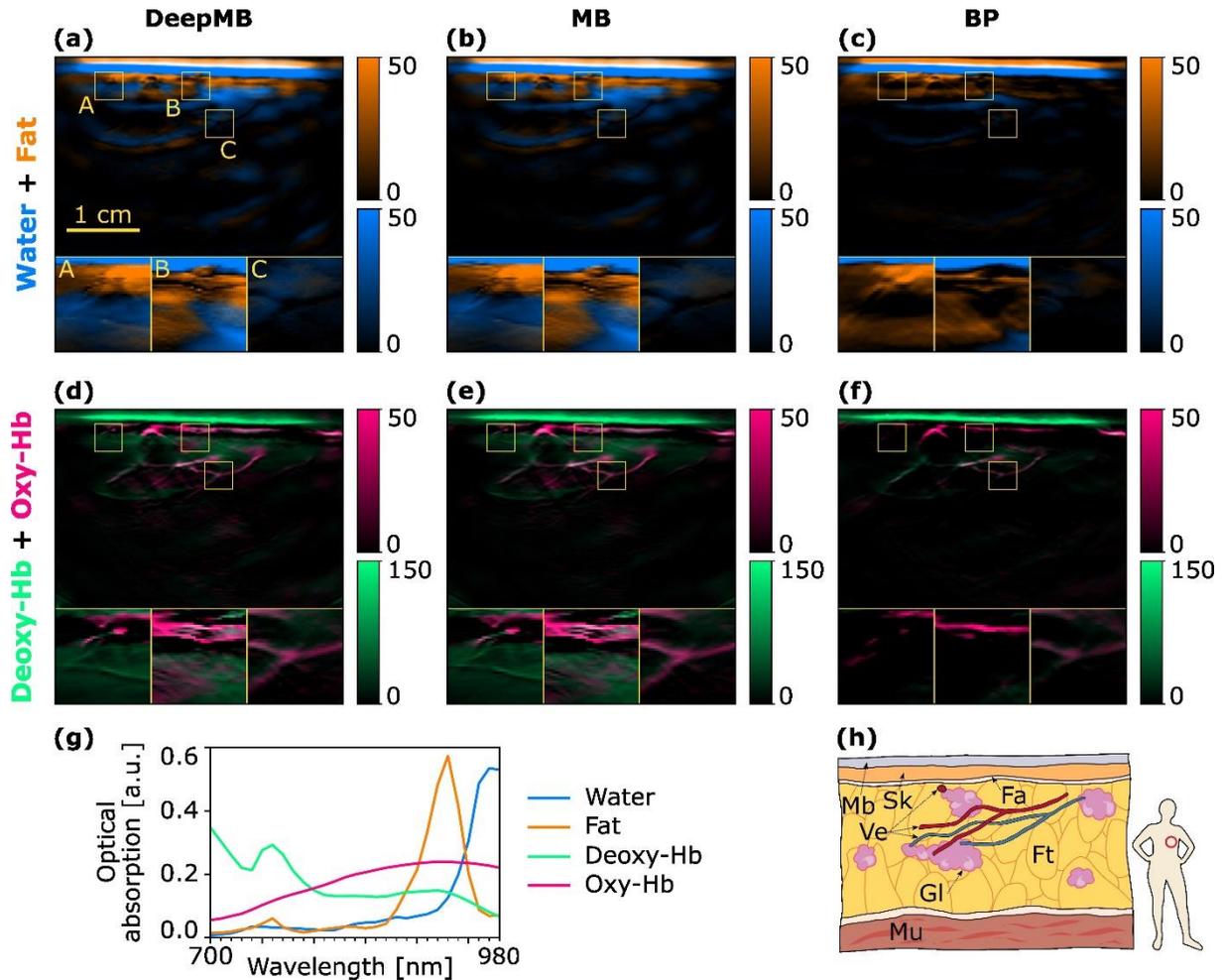

Figure 4: Unmixing of a representative multispectral breast scan for deep model-based (DeepMB; a, d), model-based (MB; b, e), and backprojection (BP; c, f). The unmixed components for fat and water and for oxyhemoglobin and deoxyhemoglobin are shown in the first two rows, respectively. The third row depicts the reference absorption spectra of the four chromophores used during unmixing (g) and a schematic sketch of the anatomical context for the depicted scan (h). All optoacoustic images show the unmixed components in arbitrary units and were slightly cropped to a field of view of 4.16×2.80 cm$^2$ to disregard the area occupied by the probe couplant above the skin line. Mb: probe membrane, Sk: skin, Fa: fascia, Mu: muscle, Ve: blood vessel, Ft: fat, Gl: glandular tissues.

## Comparison with alternative training strategies for DeepMB

The evaluation experiments described so far thoroughly validate the ability of DeepMB to reconstruct high-quality images with adjustable SoS values from the range 1475–1525 m/s. Additionally, alternative training strategies were assessed to better understand the effects of different specific aspects of the DeepMB methodology on the obtained image quality (more precisely, the SoS encoding scheme, the choice of reference training images, and the type of training data).

**Advantages of one-hot-encoded SoS values.** Passing the one-hot-encoded input SoS value to the trainable layers of the network (as shown in Fig. 1d) slightly improves the image fidelity (i.e., the data residual norms) of DeepMB reconstructions. To evaluate the benefits of this strategy, two other models with alternative SoS encoding schemes were trained and assessed: the first without providing the SoS to the U-Net (referred to as DeepMB$_{no-sos}$), and the second with the SoS encoded as a scalar value into one additional input channel for the U-Net (referred to as DeepMB$_{scalar-sos}$). The SoS was in both models used to apply the delay operator before the trainable U-Net layers, analogously to the standard DeepMB model (see Fig. 1d). Not providing the SoS as input to the U-Net was found to be a marginally inferior alternative to the standard one-hot-based SoS encoding with respect to image fidelity: DeepMB$_{no-sos}$ inferred high-quality and artifact-free images with visually the same quality as the standard DeepMB model, however with in average slightly higher data residual norms (0.164 vs. 0.156, also see Table 1). Further quantitative comparison of DeepMB and DeepMB$_{no-sos}$ reconstructions with image-based metrics did not identify a clearly superior approach (also see Table 1), which corroborates that their overall visual appearance is very similar. Providing the SoS as scalar value to the U-Net was found to be a disadvantageous encoding scheme that impedes the ability of the neural network to learn an accurate reconstruction operator, because the overall brightness of images reconstructed with $DeepMB_{scalar-sos}$ was found to be associated with the input SoS values. More specifically, inferring $DeepMB_{scalar-sos}$ onto the same sinogram with different input SoS values obtained images of higher average intensities for higher input SoS values. These intensity differences were visually imperceptible with default colormaps but resulted in notably higher average data residual norms for the obtained images in comparison to $DeepMB_{no-sos}$ or the standard DeepMB model (0.169 vs. 0.164 or 0.156, also see Table 1 for additional evaluation metrics).

**Advantages of model-based reference images.** Using model-based reference images as ground-truth references during training is essential to learn a generalizable model-based reconstruction operator. To compare the training strategy of DeepMB to the training methodology reported in previous deep-learning-based reconstruction methods for which the learning reference was true initial pressure images[29-34], another alternative model, referred to as $DeepMB_{initial-images}$, was trained using as ground-truth references the true synthetic initial pressure images (left side of Fig. 1a) instead of model-based reconstructions (right side of Fig. 1c). The reconstruction operator learnt by $DeepMB_{initial-images}$ was inferior in comparison to the standard DeepMB model: In vivo images reconstructed with $DeepMB_{initial-images}$ suffer from low resolution and contrast (see Extended Data Fig. 6) and have notably worse data residual norms (mean±std = 0.267±0.094, also see Table 1 for additional evaluation metrics) than the standard DeepMB model.

**Advantages of synthesized training data.** Synthesized training data enables DeepMB to learn an accurate and general reconstruction operator. To contextualize the image quality of DeepMB with synthesized training data, alternative DeepMB models were trained on in vivo data instead of real-world images (see section "Methods / Network training"). These models, referred to as $DeepMB_{in-vivo}$, inferred images with, in average, slightly better data residual norms than the standard DeepMB model (0.155 vs. 0.156, also see Table 1 for additional evaluation metrics). However, approximately 20% of all $DeepMB_{in-vivo}$ images contained visible artifacts, either at the left or right image borders, or in regions showing strong absorption at the skin surface. Extended Data Fig. 7 shows representative examples of such artifacts. No artifacts were observed with the standard DeepMB model (trained using synthesized data), even when reducing the size of the synthetic training set from 8000 to 3500 to match the reduced amount of available in vivo training data.

|  |  | Our method | Reference method | Traditional methods | | Alternative DeepMB training strategies | | | |
| --- | --- | --- | --- | --- | --- | --- | --- | --- | --- |
|  |  | DeepMB | MB | BP | DMAS-CF | DeepMB$_{no\text{-}sos}$ | DeepMB$_{scalar\text{-}sos}$* | DeepMB$_{initial\text{-}images}$** | DeepMB$_{in\text{-}vivo}$*** |
| In-focus images | R (↓) | 0.156 [0.092, 0.189] | 0.139 [0.068, 0.180] | 0.369 [0.294, 0.428] | 0.982 [0.972, 0.996] | 0.164 [0.106, 0.193] | 0.169 [0.113, 0.197] | 0.267 [0.196, 0.324] | 0.155 [0.088, 0.193] |
|  | MAE (↓) | 0.74 [0.43, 0.75] | n/a | 3.98 [2.60, 4.50] | 4.58 [2.80, 5.14] | 0.72 [0.41, 0.74] | 0.80 [0.47, 0.81] | 18.23 [13.49, 21.07] | 0.59 [0.30, 0.57] |
|  | MAE$_{rel}$ (%) (↓) | 15.21 [12.90, 17.21] | n/a | 86.42 [82.43, 90.50] | 96.60 [95.58, 98.31] | 14.81 [12.35, 16.88] | 16.54 [14.11, 18.53] | 429.55 [357.69, 494.13] | 11.42 [9.57, 12.78] |
|  | MSE (↓) | 9.45 [0.56, 2.41] | n/a | 84.98 [24.97, 85.20] | 254.85 [60.57, 236.06] | 8.51 [0.56, 3.15] | 10.16 [0.58, 3.41] | 703.59 [325.74, 837.51] | 5.35 [0.43, 1.70] |
|  | MSE$_{rel}$ (%) (↓) | 1.34 [0.65, 1.49] | n/a | 37.01 [29.18, 43.85] | 94.85 [93.48, 97.46] | 1.47 [0.65, 1.84] | 1.71 [0.82, 2.00] | 455.05 [265.24, 574.32] | 0.93 [0.50, 1.03] |
|  | SSIM (↑) | 0.98 [0.98, 0.99] | n/a | 0.73 [0.68, 0.79] | 0.65 [0.61, 0.69] | 0.98 [0.98, 0.99] | 0.98 [0.97, 0.99] | 0.37 [0.31, 0.42] | 0.99 [0.99, 0.99] |
| Out-of-focus images | R (↓) | 0.166 [0.087, 0.222] | 0.149 [0.059, 0.212] | 0.365 [0.281, 0.439] | 0.982 [0.971, 0.997] | 0.176 [0.105, 0.228] | 0.181 [0.111, 0.230] | 0.275 [0.196, 0.342] | 0.164 [0.081, 0.226] |
|  | MAE (↓) | 0.78 [0.42, 0.76] | n/a | 4.10 [2.58, 4.49] | 4.72 [2.74, 5.14] | 0.77 [0.40, 0.76] | 0.83 [0.46, 0.82] | 18.43 [13.56, 20.87] | 0.61 [0.30, 0.59] |
|  | MAE$_{rel}$ (%) (↓) | 15.12 [12.37, 17.21] | n/a | 86.34 [81.90, 90.79] | 96.53 [95.45, 98.01] | 14.85 [12.29, 17.29] | 16.49 [14.08, 18.55] | 425.41 [356.87, 486.35] | 11.42 [9.67, 12.81] |
|  | MSE (↓) | 13.85 [0.51, 2.56] | n/a | 92.27 [24.21, 85.79] | 295.52 [52.93, 240.66] | 11.87 [0.52, 3.64] | 13.79 [0.63, 3.64] | 711.53 [320.39, 783.03] | 6.99 [0.39, 1.82] |
|  | MSE$_{rel}$ (%) (↓) | 1.41 [0.65, 1.54] | n/a | 36.78 [28.22, 45.58] | 94.54 [93.25, 97.13] | 1.55 [0.64, 2.04] | 1.80 [0.84, 2.10] | 445.76 [259.31, 570.82] | 0.89 [0.52, 1.01] |
|  | SSIM (↑) | 0.98 [0.97, 0.98] | n/a | 0.71 [0.65, 0.79] | 0.62 [0.58, 0.67] | 0.98 [0.98, 0.99] | 0.98 [0.97, 0.98] | 0.36 [0.31, 0.41] | 0.99 [0.99, 0.99] |

*All DeepMB$_{scalar\text{-}sos}$ images systematically have their overall brightness associated with the input SoS. **All DeepMB$_{initial\text{-}images}$ images suffer from strong reconstruction artifacts that manifest as intensity saturation (see Extended Data Fig. 6). ***Some DeepMB$_{in\text{-}vivo}$ images suffer from visible reconstruction artifacts that manifest as coffee-stain-like structures (see Extended Data Fig. 7).

Table 1. Quantitative evaluation of the image quality for all reconstruction methods assessed in this paper using the data residual norm (R), the mean absolute error (MAE and MAE$_{rel}$), the mean squared error (MSE and MSE$_{rel}$), and the structural similarity index (SSIM), in comparison to the reference model-based reconstruction. The table shows the mean values and in square brackets the 25$^{th}$ and 75$^{th}$ percentiles for in-focus images (4814 in vivo sinograms from the test dataset reconstructed with each one optimal SoS values) and out-of-focus images (638 in vivo sinograms from the test dataset reconstructed each with all 11 available SoS values). The arrow symbols (↑) and (↓) indicate for each metric whether a higher or lower value is better. SoS: speed of sound. DeepMB: deep model-based. MB: model-based. BP: backprojection. DMAS-CF: delay-multiply-and-sum with coherence factor. DeepMB$_{no\text{-}sos}$: training conducted without providing the SoS as additional input to the U-Net. DeepMB$_{scalar\text{-}sos}$: training conducted with encoding the SoS value into one additional input channel for the U-Net. DeepMB$_{initial\text{-}images}$: training conducted on the true synthetic initial pressure images instead of the corresponding MB reconstructions. DeepMB$_{in\text{-}vivo}$: training conducted on in vivo data instead of synthetic data.

|  | Our method | Traditional methods | | Alternative DeepMB training strategies | | | |
|---|---|---|---|---|---|---|---|
|  | DeepMB | BP | DMAS-CF | DeepMB$_{no\text{-}sos}$ | DeepMB$_{scalar\text{-}sos}$ | DeepMB$_{initial\text{-}images}$ | DeepMB$_{in\text{-}vivo}$ |
| MAE ($\downarrow$) | 1.26 [0.80, 1.50] | 5.34 [3.91, 6.04] | 7.78 [5.82, 8.78] | 1.26 [0.77, 1.54] | 1.39 [0.90, 1.61] | 29.22 [24.58, 32.07] | 1.06 [0.62, 1.20] |
| MAE$_{rel}$ (%) ($\downarrow$) | 15.34 [13.20, 16.83] | 67.46 [65.47, 69.80] | 98.90 [98.39, 99.71] | 15.26 [13.10, 16.70] | 17.01 [15.07, 18.47] | 390.97 [342.72, 431.73] | 12.51 [10.50, 13.57] |
| MSE ($\downarrow$) | 52.18 [3.52, 38.8] | 337.88 [96.38, 423.92] | 1527.62 [476.96, 1666.41] | 51.38 [4.05, 46.39] | 62.23 [4.67, 49.09] | 3344.46 [1704.83, 4146.30] | 31.07 [3.10, 19.14] |
| MSE$_{rel}$ (%) ($\downarrow$) | 1.55 [0.72, 1.91] | 21.11 [18.21, 23.85] | 97.40 [95.93, 98.39] | 1.78 [0.86, 2.21] | 2.16 [1.02, 2.82] | 295.44 [199.81, 363.65] | 1.06 [0.60, 1.06] |
| SSIM ($\uparrow$) | 0.99 [0.99, 1.00] | 0.90 [0.87, 0.93] | 0.83 [0.80, 0.87] | 0.99 [0.99, 1.00] | 0.99 [0.99, 1.00] | 0.59 [0.50, 0.70] | 1.00 [1.00, 1.00] |

Table 2. Quantitative comparison of the unmixing components from DeepMB, BP, and all alternative DeepMB models with the unmixing components from reference model-based reconstruction using the mean absolute error (MAE and MAE$_{rel}$), the mean squared error (MSE and MSE$_{rel}$), and the structural similarity index (SSIM). The table shows the mean values and in square brackets the 25$^{th}$ and 75$^{th}$ percentiles for the 166 multispectral stacks from the in vivo test dataset. The arrow symbols ($\uparrow$) and ($\downarrow$) indicate for each metric whether a higher or lower value is better. DeepMB: deep model-based. BP: backprojection. DMAS-CF: delay-multiply-and-sum with coherence factor. DeepMB$_{no\text{-}sos}$: training conducted without providing the SoS as additional input to the U-Net. DeepMB$_{scalar\text{-}sos}$: training conducted with encoding the SoS value into one additional input channel for the U-Net. DeepMB$_{initial\text{-}images}$: training conducted on the true synthetic initial pressure images instead of the corresponding MB reconstructions. DeepMB$_{in\text{-}vivo}$: training conducted on in vivo data instead of synthetic data.

## 3. Discussion

MSOT is a high-resolution functional imaging modality that can non-invasively quantify a broad range of pathophysiological phenomena by accessing the endogenous contrast of chromophores in tissue[9]. Model-based reconstruction affords optoacoustic images with state-of-the-art quality, but remains inaccessible during real-time imaging because of the iterative and computationally demanding nature of the algorithm. In this work, we introduce a deep neural network to learn the model-based reconstruction operator and infer images with qualities nearly-identical to model-based reconstruction in 31 ms per image. Our reconstruction framework, termed DeepMB, enables accurate generalization to in vivo measurements and dynamic adjustment of the reconstruction SoS during imaging. Furthermore, DeepMB is designed to be compatible with the data rates and image sizes of modern MSOT scanners. DeepMB can therefore enable dynamic-imaging applications of optoacoustic tomography and deliver high-quality images to clinicians in real-time during examinations, furthering the clinical translation of this technology and leading to more accurate diagnoses and surgical guidance.

We trained DeepMB on synthesized sinograms from real-world images instead of in vivo images because these synthesized sinograms afford a large training dataset with a versatile set of image features, allowing DeepMB to accurately reconstruct images with diverse features. In particular, such general-feature training datasets reduce the risk of encountering out-of-distribution samples (test data with features that are not contained in the training dataset) when applying the trained model to in vivo scans. In contrast, training a model on in vivo scans systematically introduces the risk of overfitting to specific characteristics of the training samples and could potentially lead to decreased image quality for never-seen-before scans that may involve different anatomical views or disease states. We indeed observed that alternative DeepMB$_{in\text{-}vivo}$ models trained on in vivo data failed to adequately generalize to some in vivo test scans and introduced artifacts within the reconstructed images (see Extended Data Fig. 7). Furthermore, using synthesized data instead of in vivo data alleviates the training of new DeepMB models because it obviates the need for recruiting and scanning a cohort of participants. Instead, training data can be automatically generated and used to straightforwardly obtain specifically-trained DeepMB models for new scanners or different reconstruction parameters. On the other hand, our quantitative evaluation with data residual norms and image-based metrics showed that the use of more domain-specific training data (in our case in vivo scans) facilitated in aggregate slightly better images than the standard DeepMB model (e.g., average data residual norms of 0.155 for DeepMB$_{in\text{-}vivo}$ vs. 0.156 for

DeepMB, see Table 1 for all evaluation metrics). Domain-specific training data can improve the reconstruction performance because it facilitates learning of a domain-specific data transform that exploits inherent characteristics and local spatial correlation of the considered data manifold[18]. Overall, the trade-off between domain-specific training data to improve accuracy and general training data to reduce the risk of out-of-distribution samples remains a fundamental challenge for real-world application of deep learning[43, 44]. Therefore, subsequent research may focus on strategies for balancing generality, accuracy, and practicality during model training, e.g. by employing hybrid training sets combining synthesized data from real-world images with in vivo optoacoustic images and synthesized data from other biomedical scenes, or by applying domain-adaptation techniques[36, 45, 46].

Accurate generalization from synthesized training to in vivo test data is possible with DeepMB because the underlying inverse problem to solve (that is, regularized model-based reconstruction[41]) is well-posed; for each input sinogram there is a unique and stable solution (i.e., the reconstructed image). Therefore, the network can learn a data transform that is agnostic to specific characteristics of the ground-truth images during training and generalizes to images with any content (be it synthesized or in vivo)[18]. In contrast, the alternative model DeepMB$_{initial-images}$ trained on true synthetic initial pressure images (left side in Fig. 1a) falls short to accurately generalize to experimental test data and ultimately results in decreased reconstruction image quality for in vivo data (see Table 1 and Extended Data Fig. 6) because the underlying inverse problem is ill-posed. More specifically, true synthetic initial pressure images contain information not available in the input sinograms due to limited angle acquisition, measurement noise, and finite transducer bandwidth. To restore the missing information, DeepMB$_{initial-images}$ must incorporate information from the training data manifold, which hinders the correct processing of test data not contained in the training data manifold.

DeepMB supports dynamic adjustments of the SoS parameter during imaging to reconstruct high-resolution and in-focus images for arbitrary tissue types. Information about the SoS in the imaged region is required during optoacoustic image reconstruction to compute the travel time of acoustic signals between the source chromophores and the transducers of the imaging system, and to account for the spatial impulse response of the imaging system.[14, 15] In practice, the optimal SoS for a reconstruction is a priori unknown and needs to be manually tuned during imaging. Following previous efforts to automatically correct for SoS-related aberrations, especially in heterogeneous media[47], future research may also aim at automatically inferring the optimal SoS from the optoacoustic input sinogram — either in a distinct antecedent step or directly within the deep-learning-based reconstruction — to further improve the usability of optoacoustic imaging.

The presented methodology to accelerate iterative model-based reconstruction is also applicable to other optoacoustic reconstruction approaches. For instance, frequency-band model-based reconstruction[48] or Bayesian optoacoustic reconstruction[49, 50] can disentangle structures of different physical scales and quantifying reconstruction uncertainty, respectively, but their long reconstruction times currently hinder their use in real-time applications. The underlying methodology of DeepMB could also be exploited to accelerate parametrized (iterative) inversion approaches for other imaging modalities, such as ultrasound[51], X-ray computed tomography[17, 52], magnetic resonance imaging[26-28, 53], computed tomography[25], or, more generally, for any parametric partial differential equation[24]. In conclusion, we introduced DeepMB as a fully operational software-based prototype for real-time model-based optoacoustic image reconstruction. We are currently working on embedding DeepMB into the hardware of a modern MSOT scanner, to use DeepMB for real-time imaging in clinical applications.

## 4. Methods

### Handheld MSOT imaging system

We evaluated DeepMB with a modern handheld MSOT scanner (MSOT Acuity Echo, iThera Medical GmbH, Munich, Germany). The system is equipped with a multi-wavelength laser that illuminates tissues with short laser pulses (<10 ns) at a repetition rate of 25 Hz. The scanner features a custom-made ultrasound detector (IMASONIC SAS, Voray-sur-l'Ognon, France) with the following characteristics: Number of piezoelectric elements: 256; Concavity radius: 4 cm; Angular coverage: 125°; Central frequency: 4 MHz. Parasitic noise generated by light-transducer interference is reduced via optical shielding of the matching layer, yielding an extended 153% frequency bandwidth. The raw channel data for each optoacoustic scan is recorded with a sampling frequency of 40 MHz during 50.75 µs, yielding a sinogram of size 2030×256 samples. Co-registered B-mode ultrasound images are also acquired and interleaved at approximately 6 Hz for live guidance and navigation. During imaging, optoacoustic backprojection images as well as B-mode ultrasound images are displayed in real time on the scanner monitor for guidance.

**Acquisition of in vivo test sinograms**

To collect in vivo data for DeepMB evaluation, we scanned six healthy volunteers. The involved participants were three females and three males, aged from 20 to 36 years (mean age: 28.3±5.7). Self-assessed skin color according to the Fitzpatrick scale was type II (2 participants), type III (3 p.), and type IV (1 p.). Self-assessed body type was ectomorph (2 p.), mesomorph (3 p.), and endomorph (1 p.). We have complied with all relevant ethical regulations following the guidelines provided by Helmholtz Center Munich. All participants gave written informed consent upon recruitment.

For each participant, we scanned between 25 and 29 different combinations of anatomical locations and probe orientations: biceps, thyroid, carotid, calf (each left/right and transversal/longitudinal), elbow, neck, colon (each left/right), and breast (each left/right and top/bottom, female participants only). For each combination of anatomical location and probe orientation, we conducted between one and four acquisitions. During each acquisition, we recorded sinograms for approximately 10 s at wavelengths cyclically iterating from 700 to 980 nm in steps of 10 nm. We then selected, per acquisition, the 29 consecutively acquired sinograms for which we observed minimal motion in the interleaved ultrasound images, amounting to a total of 4814 in vivo test sinograms.

Finally, we band-pass filtered all selected in vivo sinograms between 100 kHz and 12 MHz to remove frequency components beyond the transducer bandwidth and cropped the first 110 time samples to remove device-specific noise present at the beginning of the sinograms.

**Determination of the SoS values**

To evaluate DeepMB reconstructions under both in-focus and out-of-focus conditions, we manually tuned the SoS value of all in vivo test scans. We used a SoS step size of 5 m/s to enable SoS adjustments slightly below the system spatial resolution (approximatively 200 µm). We found that the range of optimal SoS values was 1475–1525 m/s for the in vivo dataset, and we therefore used the same range to define the supported input SoS values of the DeepMB network.

For each scan, we manually selected the SoS value that resulted in the most well-focused reconstructed image. To speed up tuning, we selected the optimal SoS values based on approximate and high-frequency-dominated reconstructions that we computed by applying the transpose model of the system to the recorded sinograms. Furthermore, we tuned the SoS only for scans at 800 nm and adopted the values for all scans at other wavelengths acquired at the time exploiting their spatial co-registration due to the absence of motion (see previous sections for details).

**Synthesis of sinograms for training and validation**

For network training and validation, optoacoustic sinograms were synthesized with an accurate physical forward model of imaging process that incorporates the total impulse response of the system[14], parametrized by a SoS value drawn uniformly at random from the range 1475–1525 m/s with step size 5 m/s. Real-world images serving as initial pressure distributions for the forward simulations were randomly selected from the publicly available PASCAL Visual Object Classes Challenge 2012 (VOC2012) dataset[37] converted to mono-channel grayscale, and resized to 416×416 pixels. After the application of the forward model, each synthesized sinogram was scaled by a factor drawn uniformly at random from the range 0–450 to better match the variance observed in in vivo sinograms.

**Image reconstruction**

To generate ground-truth optoacoustic images, we reconstructed all sinograms (synthetic as well as in vivo) via iterative-model-based. We used Shearlet $L^1$ to tackle the ill-posedness of the inverse problem. Shearlet $L^1$ regularization is a convex relaxation of Shearlet sparsity, which can reduce limited-view artifacts in reconstructed images, because Shearlets provide a maximally-sparse approximation of a larger class of images (known as cartoon-like functions) with a mathematically-proven optimal encoding rate[54]. The optimal pressure field to find is characterized as

$$p_0 := \arg\min_{p\geq 0} \|M_{SoS}\, p - s\|_2^2 + \lambda\|SH(p)\|_1,$$

where $p_0$ is the reconstructed image, $M_{SoS}$ is the forward model of the imaging process for the selected reconstruction SoS, s is the input sinogram, $\lambda$ is the regularization parameter tuned via an L-curve, SH is the Shearlet transform, and $\|\cdot\|_n$ is the n-norm. The minimization problem was solved via bound-constrained sparse reconstruction by separable approximation[55-57]. All images were reconstructed with a size of 416×416 pixels and a field of view of 4.16×4.16 cm$^2$. For comparison purposes, we also reconstructed all images using the backprojection formula[58, 59] and the delay-multiply-and-sum with coherence factor algorithm[39, 40].

**Network training**

The DeepMB network was implemented in Python and PyTorch. It was trained — either on synthetic or in vivo data — for 300 epochs using stochastic gradient descent with batch size=4, learning rate=0.01, momentum=0.99, and per epoch learning rate decay factor=0.99. The network loss was calculated as the mean square error between the output image and the reference image. The final model was selected based on the minimal loss on the validation dataset, and finally compiled into an ONNX model for speed-up.

To facilitate training, all input sinograms were scaled by K=450$^{-1}$ to ensure that their values never exceed the range [-1, 1]. The same scaling factor was also applied to all target images. Furthermore, the square root was applied to all target reference images used during training and validation to reduce the network output values and limit the influence of high intensity pixels during loss calculation. When applying the trained network on in vivo test data, inferred images were first squared then scaled by K$^{-1}$, to revert the preprocessing operation.

When training on synthetic data to build the standard DeepMB model, we used 8000 sinograms as train split and 2000 sinograms as validation split. The alternative scenario involving training on in vivo data to build the DeepMB$_{in-vivo}$ models was carried out as described hereafter: six different permutations were conducted, with a 4/1/1 participants division between the train, validation, and test splits, respectively, each participant being once and only once part of the validation and test splits.

The DeepMB network is based upon the U-Net architecture[38] with a depth of 5 layers and a width of 64 features. To gradually reduce the total number of data channels from 267 (that is, 256 transducer elements, and one-hot encoding of 11 possible SoS values) down to 64, three 2D convolutional layers with 208, 160, and 112 features, respectively, were added prior to the U-Net. All kernel and padding size were (3, 3) and (1, 1), respectively. Biases were accounted for, and the final activation was the absolute value function.

**Data residual norm**

To quantify the image fidelity of reconstructions from DeepMB, model-based, or backprojection, we evaluated the data residual norm R, defined as

$$R := \frac{\|M_{SoS}\, p_0 - s\|_2^2}{\|s\|_2^2},$$

where $p_0$ is the reconstructed image, $M_{SoS}$ is the forward model from model-based reconstruction, s is the input sinogram, and $\|\cdot\|_2$ is the 2-norm. Time sample values from the input sinogram that are outside the reach of the applied forward model are set to zero prior to computing the data residual norm to avoid distortions by signals originating from outside the field of view. We employed data residual norms as the primary evaluation metric for our experiments because it respects the underlying physics of the imaging process and is provably minimal for model-based reconstruction. To constrain the solutions space for all reconstruction methods in a similar way and enable a meaningful comparison between backprojection on one hand, versus non-negative model-based and DeepMB on the other hand, negative pixel values were set to zero prior to residual calculation for backprojection images. All images were individually scaled using the linear degree of freedom in reconstructed optoacoustic image so that their data residual norms are minimal.

For the evaluation of in-focus images, data residual norms were calculated for the reconstructions with the optimal SoS values of all 4814 samples from the in vivo test set. For the evaluation of out-of-focus images, data residuals were calculated for the reconstructions with all 11 SoS values of a subset of 638 randomly selected in vivo samples.

**Unmixing**

To evaluate the multispectral image quality of DeepMB, model-based, and backprojection, all reconstructed in-vivo scans from the test dataset were grouped into multispectral stacks of 29 images (respectively one scan from the range 700–980 nm in steps of 10 nm) and unmixed into oxyhemoglobin, deoxyhemoglobin, fat, and water components:

$$\widehat{W} := \arg\min_{W \geq 0} \|S - WH\|_F^2,$$

where S (size 173056×29) denotes all pixels of a multispectral stack, H (size 4×29) denotes the reference absorption spectra of water, fat, oxyhemoglobin, and deoxyhemoglobin in the wavelength range 700–980 nm, and $\widehat{W}$ (size 173056×4) denotes the unmixed components for the four considered chromophores. $\|M\|_F := \left(\sum_{i,j} m_{i,j}^2\right)^{0.5}$ denotes the Frobenius norm and $M \geq 0$ refers to entry wise inequality. All negative pixel values in the backprojection images were set to zero prior to unmixing.

**Image-based evaluation metrics**

Additionally, we quantified the deviation of standard DeepMB, all alternative DeepMB, and backprojection images from reference model-based reconstructions by computing the mean absolute error (MAE), the relative mean absolute error (MAE$_{rel}$), the mean squared error (MSE), the relative mean squared error (MSE$_{rel}$), and the structural similarity index (SSIM), defined as

$$MAE := \|i_{rec} - i_{mb}\|_1,$$

$$MAE_{rel} := \frac{\|i_{rec} - i_{mb}\|_1}{\|i_{mb}\|_1},$$

$$MSE := \|i_{rec} - i_{mb}\|_2^2,$$

$$MSE_{rel} := \frac{\|i_{rec} - i_{mb}\|_2^2}{\|i_{mb}\|_2^2},$$

$$SSIM := \frac{(2\mu_{rec}\mu_{mb} + c_1)(2\sigma_{rec,mb} + c_2)}{(\mu_{rec}^2 + \mu_{mb}^2 + c_1)(\sigma_{rec}^2 + \sigma_{mb}^2 + c_2)},$$

where $i_{rec}$ (size 173056×1) is the vectorization of a reconstructed image from either standard DeepMB, any alternative DeepMB, or backprojection, and $i_{mb}$ (size 173056×1) is the vectorization of the corresponding reference image from model-based reconstruction. SSIM is calculated as the average over sliding windows of size 21×21 pixels, where $\mu_{rec}$ and $\mu_{mb}$ are the averages of $i_{rec}$ and $i_{mb}$, $\sigma_{rec}^2$ and $\sigma_{mb}^2$ are the variances of $i_{rec}$ and $i_{mb}$, $\sigma_{rec,mb}$ is the covariance of $i_{rec}$ and $i_{mb}$, and $c_1 = (0.01 \max(i_{mb}))^2$ and $c_2 = (0.03 \max(i_{mb}))^2$ are two empirical variables to stabilize the division with weak denominators. All backprojection images were additionally preprocessed to enable a meaningful comparison with the model-based reference images: Negative pixels were set to zero and all images were individually scaled using the linear degree of freedom in reconstructed optoacoustic images so that the respectively calculated metric is minimal.

Image-based metrics were computed analogously to the data residual norms using all 4814 in vivo test samples (each reconstructed with the optimal SoS value) for the in-focus case and a subset of 638 in vivo test samples (each reconstructed with all 11 available SoS values) for the out-of-focus case.

**Code availability.** The entirety of the DeepMB source code is publicly available on GitHub[1].

---

[1] https://github.com/juestellab/deepmb

**Data availability.** The data that support the findings of this study are available on request from the corresponding author Dominik Jüstel. The data are not publicly available due to them containing information that could compromise research participant privacy/consent.

**Acknowledgements.** The authors would like to thank Antonia Longo for her precious contribution during in vivo image acquisition and the conception of Fig. 2, and Robert Wilson for his attentive reading and improvements of the manuscript.

**Funding.** This project has received funding from the Bavarian Ministry of Economic Affairs, Energy and Technology (StMWi) (DIE-2106-0005// DIE0161/02, DeepOpus) and from the European Research Council (ERC) under the European Union's Horizon 2020 research and innovation programme under grant agreement No 694968 (PREMSOT).

## 5. References

1. Ntziachristos, V. & Razansky, D. Molecular imaging by means of multispectral optoacoustic tomography (MSOT). *Chem Rev* 110, 2783-2794 (2010).
2. Diot, G., Metz, S., Noske, A., Liapis, E., Schroeder, B., Ovsepian, S.V., Meier, R., Rummeny, E. & Ntziachristos, V. Multispectral Optoacoustic Tomography (MSOT) of Human Breast Cancer. *Clin Cancer Res* 23, 6912-6922 (2017).
3. Knieling, F., Neufert, C., Hartmann, A., Claussen, J., Urich, A., Egger, C., Vetter, M., Fischer, S., Pfeifer, L., Hagel, A., Kielisch, C., Gortz, R.S., Wildner, D., Engel, M., Rother, J., Uter, W., Siebler, J., Atreya, R., Rascher, W., Strobel, D., Neurath, M.F. & Waldner, M.J. Multispectral Optoacoustic Tomography for Assessment of Crohn's Disease Activity. *N Engl J Med* 376, 1292-1294 (2017).
4. Karlas, A., Kallmayer, M., Fasoula, N.-A., Liapis, E., Bariotakis, M., Krönke, M., Anastasopoulou, M., Reber, J., Eckstein, H.-H. & Ntziachristos, V. Multispectral optoacoustic tomography of muscle perfusion and oxygenation under arterial and venous occlusion: A human pilot study. *Journal of Biophotonics* 13, e201960169 (2020).
5. Dehner, C., Olefir, I., Chowdhury, K.B., Jüstel, D. & Ntziachristos, V. Deep-learning-based electrical noise removal enables high spectral optoacoustic contrast in deep tissue. *IEEE Transactions on Medical Imaging* (2022).
6. Kukacka, J., Metz, S., Dehner, C., Muckenhuber, A., Paul-Yuan, K., Karlas, A., Fallenberg, E.M., Rummeny, E., Justel, D. & Ntziachristos, V. Image processing improvements afford second-generation handheld optoacoustic imaging of breast cancer patients. *Photoacoustics* 26, 100343 (2022).
7. Regensburger, A.P., Fonteyne, L.M., Jungert, J., Wagner, A.L., Gerhalter, T., Nagel, A.M., Heiss, R., Flenkenthaler, F., Qurashi, M., Neurath, M.F., Klymiuk, N., Kemter, E., Frohlich, T., Uder, M., Woelfle, J., Rascher, W., Trollmann, R., Wolf, E., Waldner, M.J. & Knieling, F. Detection of collagens by multispectral optoacoustic tomography as an imaging biomarker for Duchenne muscular dystrophy. *Nat Med* 25, 1905-1915 (2019).
8. Dima, A. & Ntziachristos, V. Non-invasive carotid imaging using optoacoustic tomography. *Opt Express* 20, 25044-25057 (2012).
9. Taruttis, A. & Ntziachristos, V. Advances in real-time multispectral optoacoustic imaging and its applications. *Nature Photonics* 9, 219-227 (2015).
10. Ivankovic, I., Mercep, E., Schmedt, C.G., Dean-Ben, X.L. & Razansky, D. Real-time Volumetric Assessment of the Human Carotid Artery: Handheld Multispectral Optoacoustic Tomography. *Radiology* 291, 45-50 (2019).
11. Sethuraman, S., Aglyamov, S.R., Amirian, J.H., Smalling, R.W. & Emelianov, S.Y. Intravascular photoacoustic imaging using an IVUS imaging catheter. *IEEE Trans Ultrason Ferroelectr Freq Control* 54, 978-986 (2007).
12. Yang, J.M., Maslov, K., Yang, H.C., Zhou, Q., Shung, K.K. & Wang, L.V. Photoacoustic endoscopy. *Opt. Lett.* 34, 1591-1593 (2009).
13. Xu, M. & Wang, L.V. Universal back-projection algorithm for photoacoustic computed tomography. *Phys Rev E Stat Nonlin Soft Matter Phys* 71, 016706 (2005).

14. Chowdhury, K.B., Prakash, J., Karlas, A., Jüstel, D. & Ntziachristos, V. A Synthetic Total Impulse Response Characterization Method for Correction of Hand-held Optoacoustic Images. *IEEE Transactions on Medical Imaging* 39, 3218-3230 (2020).
15. Chowdhury, K.B., Bader, M., Dehner, C., Justel, D. & Ntziachristos, V. Individual transducer impulse response characterization method to improve image quality of array-based handheld optoacoustic tomography. *Opt. Lett.* 46, 1-4 (2021).
16. Ding, L., Dean-Ben, X.L. & Razansky, D. Real-Time Model-Based Inversion in Cross-Sectional Optoacoustic Tomography. *IEEE Trans Med Imaging* 35, 1883-1891 (2016).
17. Jin, K.H., McCann, M.T., Froustey, E. & Unser, M. Deep Convolutional Neural Network for Inverse Problems in Imaging. *IEEE Transactions on Image Processing* 26, 4509-4522 (2017).
18. Zhu, B., Liu, J.Z., Cauley, S.F., Rosen, B.R. & Rosen, M.S. Image reconstruction by domain-transform manifold learning. *Nature* 555, 487-492 (2018).
19. Ongie, G., Jalal, A., Metzler, C.A., Baraniuk, R.G., Dimakis, A.G. & Willett, R. Deep Learning Techniques for Inverse Problems in Imaging. *IEEE Journal on Selected Areas in Information Theory* 1, 39-56 (2020).
20. Lucas, A., Iliadis, M., Molina, R. & Katsaggelos, A.K. Using Deep Neural Networks for Inverse Problems in Imaging: Beyond Analytical Methods. *IEEE Signal Processing Magazine* 35, 20-36 (2018).
21. Allman, D., Assis, F., Chrispin, J. & Bell, M.A.L. Deep Neural Networks to Remove Photoacoustic Reflection Artifacts in Ex Vivo and in Vivo Tissue. 2018 IEEE International Ultrasonics Symposium (IUS).
22. Gröhl, J., Schellenberg, M., Dreher, K. & Maier-Hein, L. Deep learning for biomedical photoacoustic imaging: A review. *Photoacoustics* 22, 100241 (2021).
23. Hauptmann, A. & Cox, B. Deep learning in photoacoustic tomography: current approaches and future directions. *Journal of Biomedical Optics* 25, 112903 (2020).
24. Aggarwal, H.K., Mani, M.P. & Jacob, M. MoDL: Model-Based Deep Learning Architecture for Inverse Problems. *IEEE Trans Med Imaging* 38, 394-405 (2019).
25. Liu, J., Sun, Y., Gan, W., Xu, X., Wohlberg, B. & Kamilov, U.S. SGD-Net: Efficient Model-Based Deep Learning With Theoretical Guarantees. *IEEE Transactions on Computational Imaging* 7, 598-610 (2021).
26. Genzel, M., Macdonald, J. & Marz, M. Solving Inverse Problems With Deep Neural Networks - Robustness Included. *IEEE Transactions on Pattern Analysis and Machine Intelligence*, 1-1 (2022).
27. Schlemper, J., Caballero, J., Hajnal, J.V., Price, A.N. & Rueckert, D. A Deep Cascade of Convolutional Neural Networks for Dynamic MR Image Reconstruction. *IEEE Trans Med Imaging* 37, 491-503 (2018).
28. Hammernik, K., Klatzer, T., Kobler, E., Recht, M.P., Sodickson, D.K., Pock, T. & Knoll, F. Learning a variational network for reconstruction of accelerated MRI data. *Magnetic Resonance in Medicine* 79, 3055-3071 (2018).
29. Kim, M., Jeng, G.S., Pelivanov, I. & O'Donnell, M. Deep-Learning Image Reconstruction for Real-Time Photoacoustic System. *IEEE Trans Med Imaging* 39, 3379-3390 (2020).
30. Lan, H., Jiang, D., Yang, C., Gao, F. & Gao, F. Y-Net: Hybrid deep learning image reconstruction for photoacoustic tomography in vivo. *Photoacoustics* 20, 100197 (2020).
31. Waibel, D., Gröhl, J., Isensee, F., Kirchner, T., Maier-Hein, K. & Maier-Hein, L. Reconstruction of initial pressure from limited view photoacoustic images using deep learning. Photons Plus Ultrasound: Imaging and Sensing 2018.
32. Feng, J., Deng, J., Li, Z., Sun, Z., Dou, H. & Jia, K. End-to-end Res-Unet based reconstruction algorithm for photoacoustic imaging. *Biomed. Opt. Express* 11, 5321-5340 (2020).
33. Tong, T., Huang, W., Wang, K., He, Z., Yin, L., Yang, X., Zhang, S. & Tian, J. Domain Transform Network for Photoacoustic Tomography from Limited-view and Sparsely Sampled Data. *Photoacoustics* 19, 100190 (2020).
34. Guan, S., Khan, A.A., Sikdar, S. & Chitnis, P.V. Limited-View and Sparse Photoacoustic Tomography for Neuroimaging with Deep Learning. *Sci Rep* 10, 8510 (2020).
35. Guo, M., Lan, H., Yang, C., Liu, J. & Gao, F. AS-Net: Fast Photoacoustic Reconstruction With Multi-Feature Fusion From Sparse Data. *IEEE Transactions on Computational Imaging* 8, 215-223 (2022).

36. Hauptmann, A., Lucka, F., Betcke, M., Huynh, N., Adler, J., Cox, B., Beard, P., Ourselin, S. & Arridge, S. Model-Based Learning for Accelerated, Limited-View 3-D Photoacoustic Tomography. *IEEE Trans Med Imaging* 37, 1382-1393 (2018).
37. Everingham, M., Van Gool, L., Williams, C.K.I., Winn, J. & Zisserman, A. The Pascal Visual Object Classes (VOC) Challenge. *International Journal of Computer Vision* 88, 303-338 (2010).
38. Ronneberger, O., Fischer, P. & Brox, T. U-Net: Convolutional Networks for Biomedical Image Segmentation. International Conference on Medical image computing and computer-assisted intervention 2015.
39. Jeon, S., Park, E.-Y., Choi, W., Managuli, R., Lee, K.j. & Kim, C. Real-time delay-multiply-and-sum beamforming with coherence factor for in vivo clinical photoacoustic imaging of humans. *Photoacoustics* 15, 100136 (2019).
40. Matrone, G., Savoia, A.S., Caliano, G. & Magenes, G. The Delay Multiply and Sum Beamforming Algorithm in Ultrasound B-Mode Medical Imaging. *IEEE Transactions on Medical Imaging* 34, 940-949 (2015).
41. Rosenthal, A., Ntziachristos, V. & Razansky, D. Acoustic Inversion in Optoacoustic Tomography: A Review. *Curr Med Imaging Rev* 9, 318-336 (2013).
42. Prahl, S. Assorted Spectra. https://omlc.org/spectra/ (accessed 19.01.2023).
43. Tobin, J., Fong, R., Ray, A., Schneider, J., Zaremba, W. & Abbeel, P. Domain randomization for transferring deep neural networks from simulation to the real world. 2017 IEEE/RSJ International Conference on Intelligent Robots and Systems (IROS).
44. Mårtensson, G., Ferreira, D., Granberg, T., Cavallin, L., Oppedal, K., Padovani, A., Rektorova, I., Bonanni, L., Pardini, M., Kramberger, M.G., Taylor, J.-P., Hort, J., Snædal, J., Kulisevsky, J., Blanc, F., Antonini, A., Mecocci, P., Vellas, B., Tsolaki, M., Kłoszewska, I., Soininen, H., Lovestone, S., Simmons, A., Aarsland, D. & Westman, E. The reliability of a deep learning model in clinical out-of-distribution MRI data: A multicohort study. *Medical Image Analysis* 66, 101714 (2020).
45. Susmelj, A.K., Lafci, B., Ozdemir, F., Davoudi, N., Dean-Ben, X.L., Perez-Cruz, F. & Razansky, D. Signal Domain Learning Approach for Optoacoustic Image Reconstruction from Limited View Data. Proceedings of The 5th International Conference on Medical Imaging with Deep Learning 2022.
46. Schellenberg, M., Gröhl, J., Dreher, K.K., Nölke, J.-H., Holzwarth, N., Tizabi, M.D., Seitel, A. & Maier-Hein, L. Photoacoustic image synthesis with generative adversarial networks. *Photoacoustics* 28, 100402 (2022).
47. Jeon, S., Choi, W., Park, B. & Kim, C. A Deep Learning-Based Model That Reduces Speed of Sound Aberrations for Improved In Vivo Photoacoustic Imaging. *IEEE Transactions on Image Processing* 30, 8773-8784 (2021).
48. Longo, A., Jüstel, D. & Ntziachristos, V. Disentangling the frequency content in optoacoustics. *IEEE Transactions on Medical Imaging* (2022).
49. Tick, J., Pulkkinen, A. & Tarvainen, T. Image reconstruction with uncertainty quantification in photoacoustic tomography. *J Acoust Soc Am* 139, 1951 (2016).
50. Tick, J., Pulkkinen, A., Lucka, F., Ellwood, R., Cox, B.T., Kaipio, J.P., Arridge, S.R. & Tarvainen, T. Three dimensional photoacoustic tomography in Bayesian framework. *J Acoust Soc Am* 144, 2061 (2018).
51. Hyun, D., Brickson, L.L., Looby, K.T. & Dahl, J.J. Beamforming and Speckle Reduction Using Neural Networks. *IEEE Transactions on Ultrasonics, Ferroelectrics, and Frequency Control* 66, 898-910 (2019).
52. Kang, E., Min, J. & Ye, J.C. A deep convolutional neural network using directional wavelets for low-dose X-ray CT reconstruction. *Medical Physics* 44, e360-e375 (2017).
53. Moya-Sáez, E., Peña-Nogales, Ó., Luis-García, R.d. & Alberola-López, C. A deep learning approach for synthetic MRI based on two routine sequences and training with synthetic data. *Computer Methods and Programs in Biomedicine* 210, 106371 (2021).
54. Kutyniok, G. & Lim, W.-Q. Compactly supported shearlets are optimally sparse. *Journal of Approximation Theory* 163, 1564-1589 (2011).
55. Wright, S.J., Nowak, R.D. & Figueiredo, M.A.T. Sparse Reconstruction by Separable Approximation. *IEEE Transactions on Signal Processing* 57, 2479-2493 (2009).
56. Chartrand, R. & Wohlberg, B. Total-variation regularization with bound constraints. 2010 IEEE International Conference on Acoustics, Speech and Signal Processing.

57. Kutyniok, G., Lim, W.-Q. & Reisenhofer, R. ShearLab 3D: Faithful Digital Shearlet Transforms Based on Compactly Supported Shearlets. *ACM Trans. Math. Softw.* 42, Article 5 (2016).
58. Kunyansky, L.A. Explicit inversion formulae for the spherical mean Radon transform. *Inverse Problems* 23, 373-383 (2007).
59. Kuchment, P. & Kunyansky, L. in Handbook of Mathematical Methods in Imaging. (ed. O. Scherzer) 817-865 (Springer New York, New York, NY; 2011).

## 6. Extended Data Figures

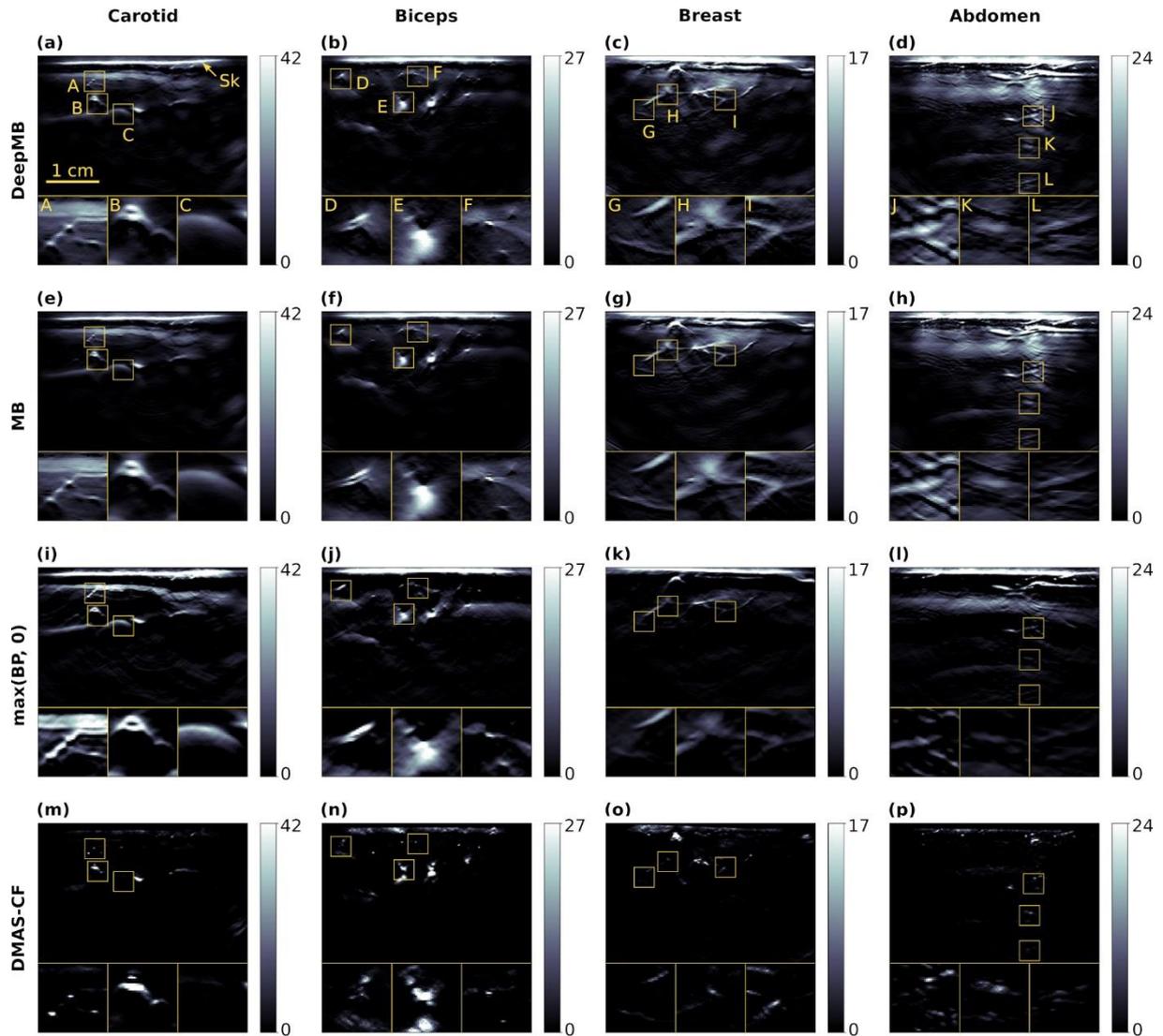

Extended Data Figure 1: Visual comparison of backprojection (BP) images with negative pixel values set to zero after the reconstruction (third row) and delay-multiply-and-sum with coherence factor (DMAS-CF, fourth row) images, against the corresponding deep model-based (DeepMB) and model-based (MB) images (first two rows). The presented samples are the same as those depicted in Figure 2. DeepMB and MB images are nearly identical; BP images notably differ from reference model-based reconstructions suffering from lower resolution (see e.g. structures shown in zoom A of tile i and zoom D of tile j), missing structures in image regions that contained negative pixel values (see e.g. zoom F of tile j, or the entire region below the skin line (Sk) in tile k and l), and reduced contrast (see e.g. structures shown in zoom I of tile k and zoom J of tile l). All images show the reconstructed initial pressure in arbitrary units and were slightly cropped to a field of view of 4.16×2.80 cm$^2$ to disregard the area occupied by the probe couplant above the skin line.

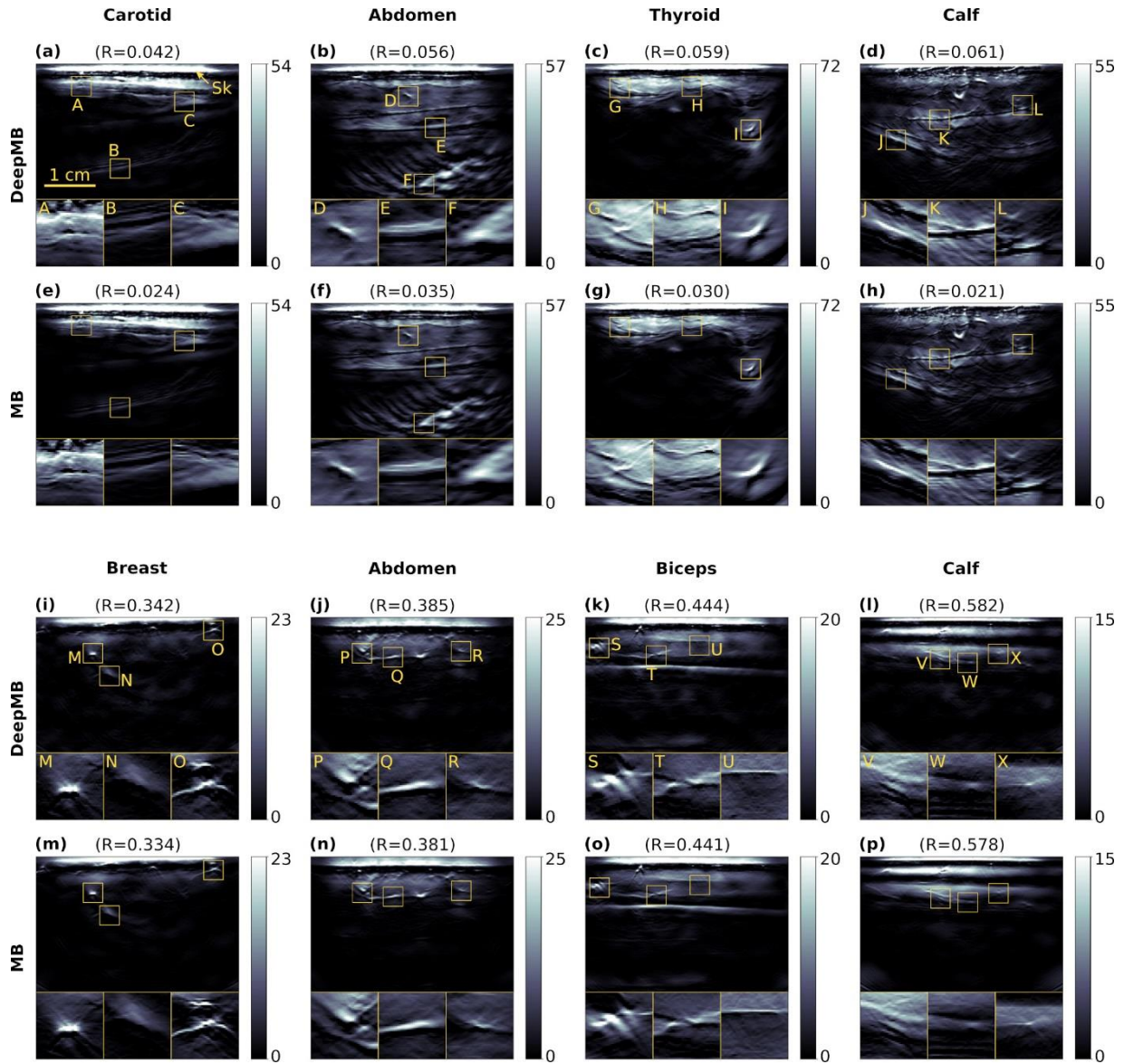

Extended Data Figure 2: Examples from the in vivo test dataset with low and high data residual norms (namely, below the 5th percentile (a-h) and above the 95th percentile (i-p) of all 4814 test samples, respectively), for deep model-based (DeepMB) and model-based (MB). The data residual norm (R) is indicated between round brackets above each image. Panels (a, e) and (l, p) correspond to the samples for which DeepMB afforded the overall lowest and highest data residual norms, respectively. All images show the reconstructed initial pressure in arbitrary units and were slightly cropped to a field of view of 4.16×2.80 $cm^2$ to disregard the area occupied by the probe couplant above the skin line (Sk).

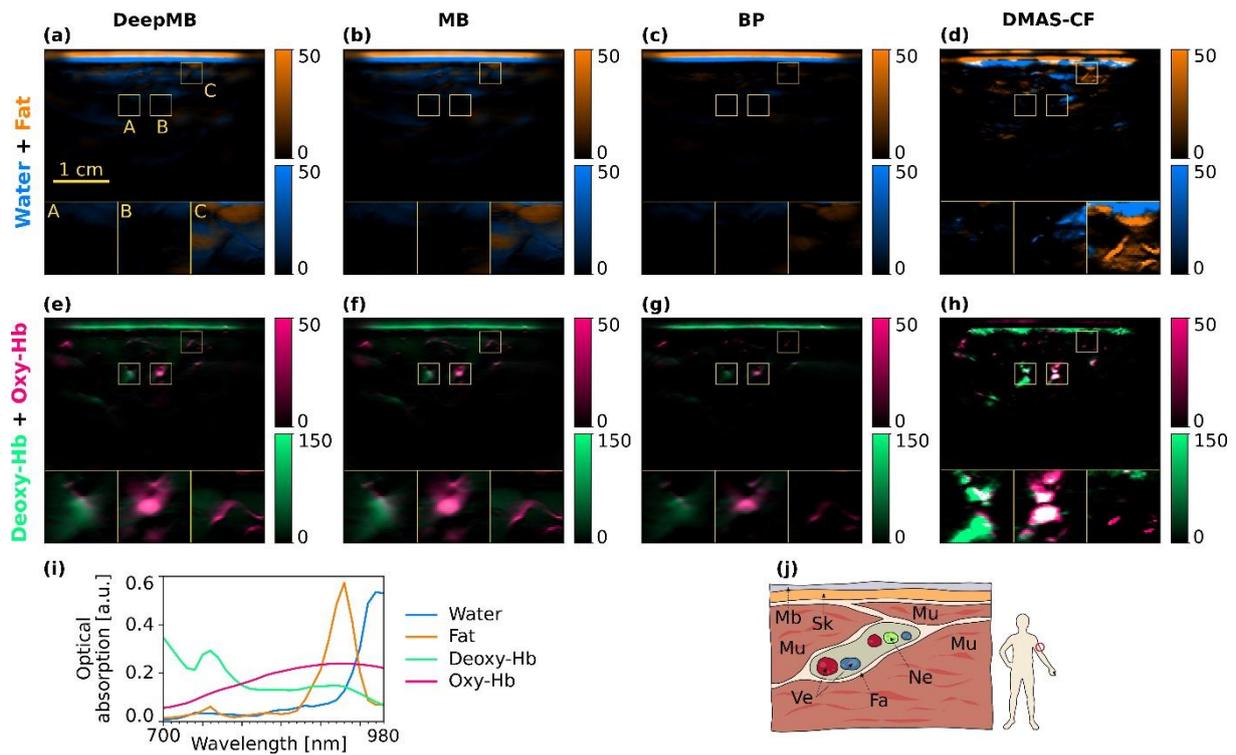

Extended Data Figure 3: Unmixing of a representative multispectral biceps scan for deep model-based (DeepMB; a, d), model-based (MB; b, e), and backprojection (BP; c, f). The unmixed components for fat and water and for oxyhemoglobin and deoxyhemoglobin are shown in the first two rows, respectively. The third row depicts the reference absorption spectra of the four chromophores used during unmixing (g) and a schematic sketch of the anatomical context for the depicted scan (h). All optoacoustic images show the unmixed components in arbitrary units and were slightly cropped to a field of view of 4.16×2.80 cm$^2$ to disregard the area occupied by the probe couplant above the skin line. Mb: probe membrane, Sk: skin, Fa: fascia, Mu: muscle, Ve: blood vessel, Ne: nerve.

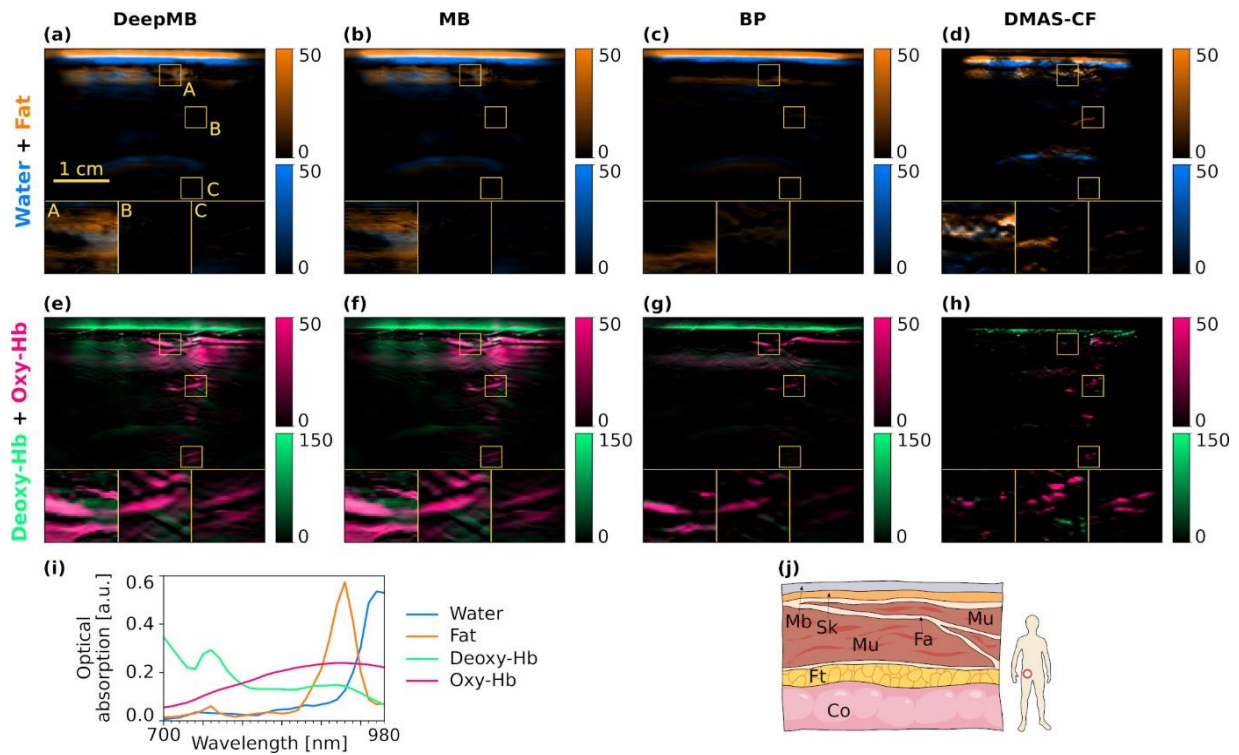

Extended Data Figure 4: Unmixing of a representative multispectral colon scan for deep model-based (DeepMB; a, d), model-based (MB; b, e), and backprojection (BP; c, f). The unmixed components for fat and water and for oxyhemoglobin and deoxyhemoglobin are shown in the first two rows, respectively. The third row depicts the reference absorption spectra of the four chromophores used during unmixing (g) and a schematic sketch of the anatomical context for the depicted scan (h). All optoacoustic images show the unmixed components in arbitrary units and were slightly cropped to a field of view of 4.16×2.80 cm$^2$ to disregard the area occupied by the probe couplant above the skin line. Mb: probe membrane, Sk: skin, Fa: fascia, Mu: muscle, Ft: fat, Co: colon.

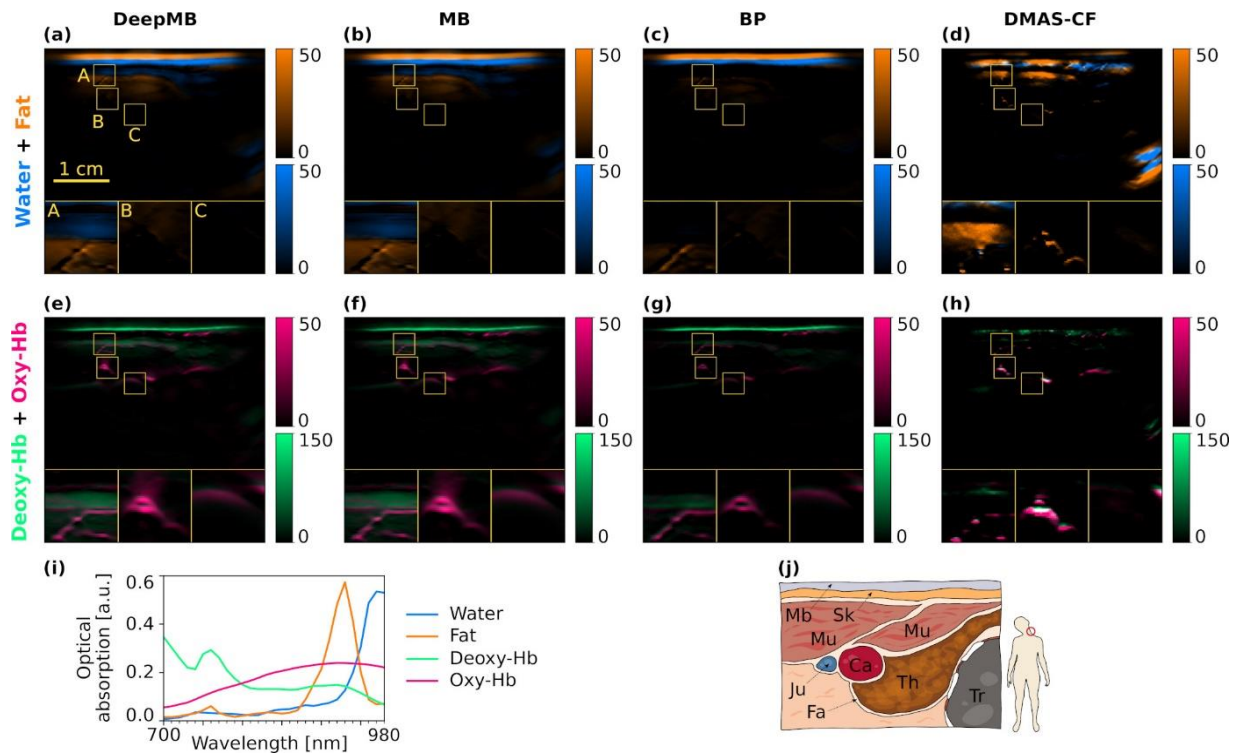

Extended Data Figure 5: Unmixing of a representative multispectral carotid scan for deep model-based (DeepMB; a, d), model-based (MB; b, e), and backprojection (BP; c, f). The unmixed components for fat and water and for oxyhemoglobin and deoxyhemoglobin are shown in the first two rows, respectively. The third row depicts the reference absorption spectra of the four chromophores used during unmixing (g) and a schematic sketch of the anatomical context for the depicted scan (h). All optoacoustic images show the unmixed components in arbitrary units and were slightly cropped to a field of view of 4.16×2.80 cm$^2$ to disregard the area occupied by the probe couplant above the skin line. Mb: probe membrane, Sk: skin, Fa: fascia, Mu: muscle, Ca: common carotid artery, Ju: jugular vein, Th: thyroid, Tr: trachea.

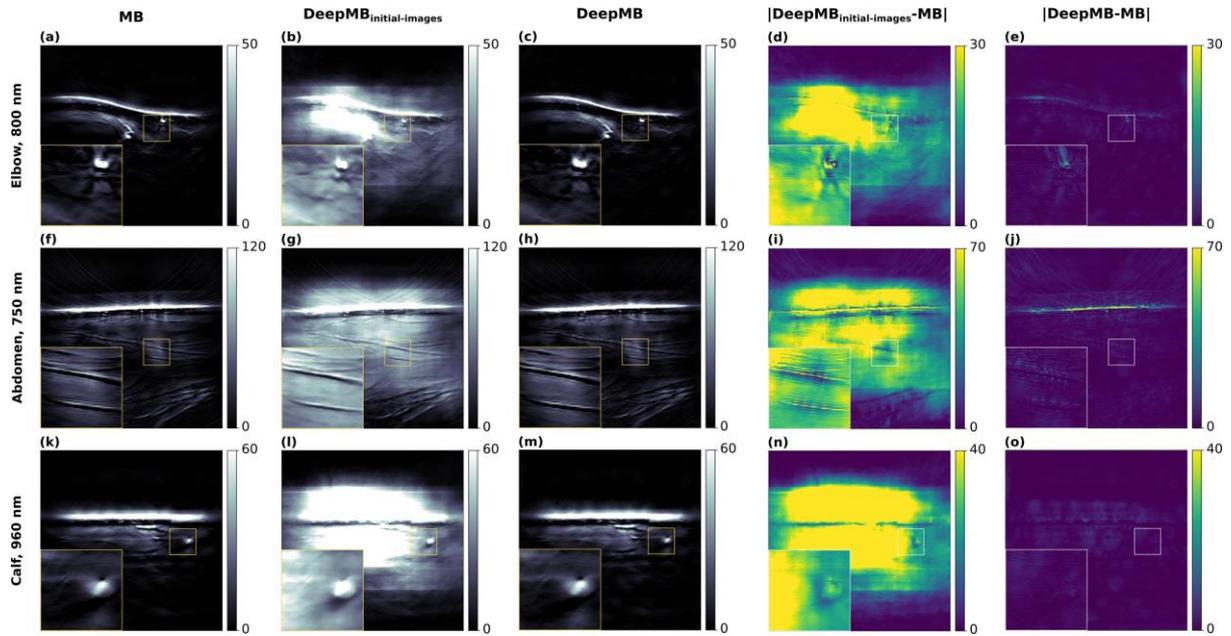

Extended Data Figure 6: Representative examples showing the inaptitude of the alternative model DeepMB$_{\text{initial-images}}$ (i.e., trained on true initial pressure images) to reconstruct in vivo images. The three rows depict different anatomies (elbow: a–e, abdomen: f–j, calf: k–o). The three leftmost columns correspond to images reconstructed via model-based (MB), alternative DeepMB$_{\text{initial-images}}$, and standard DeepMB. The two rightmost columns show the absolute differences between the reference model-based image and the image inferred from DeepMB$_{\text{initial-images}}$ and DeepMB, respectively. The field of view is 4.16×4.16 cm$^2$, the enlarged region is 0.61×0.61 cm$^2$.

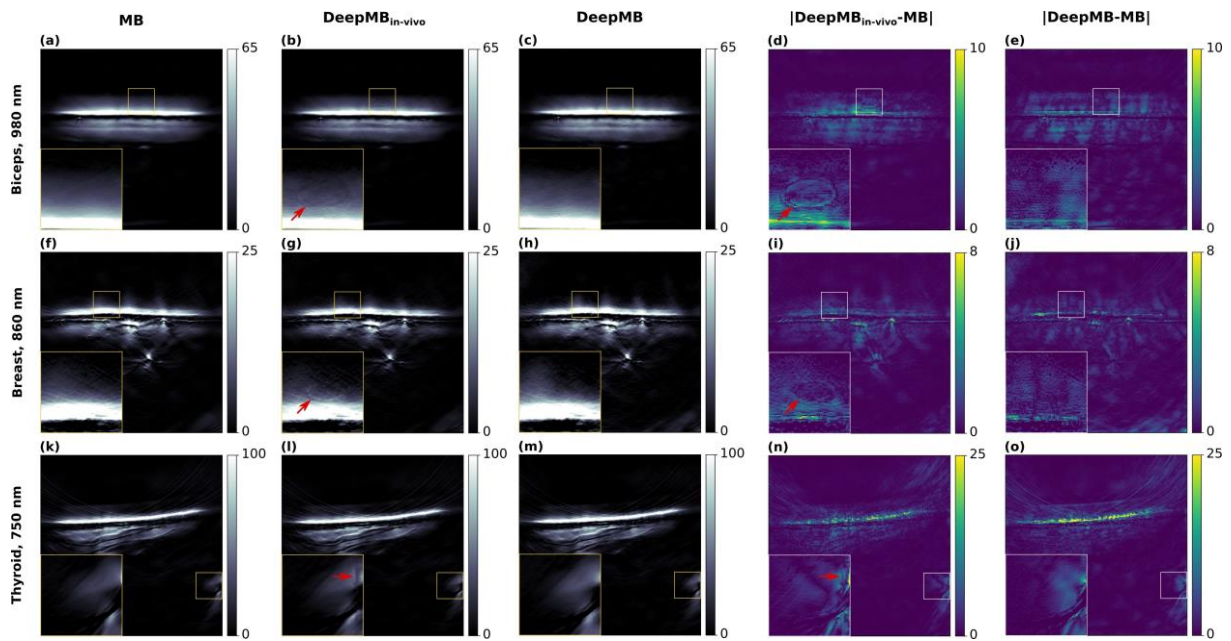

Extended Data Figure 7: Representative examples of reconstruction artifacts (red arrows) from alternative models DeepMB$_{\text{in-vivo}}$ (i.e., trained on in vivo data instead of synthesized data). The three rows depict different anatomies (biceps: a–e, breast: f–j, thyroid: k–o). The three leftmost columns correspond to images reconstructed via model-based (MB), alternative DeepMB trained on in vivo data (DeepMB$_{\text{in-vivo}}$), and standard DeepMB (DeepMB). The two rightmost columns show the absolute differences between the reference model-based image and the image inferred from DeepMB$_{\text{in-vivo}}$ and DeepMB, respectively. The field of view is 4.16×4.16 cm$^2$, the enlarged region is 0.61×0.61 cm$^2$.

# 7. Extended Data Tables

| Aggregation | Categories | DeepMB | MB |
|---|---|---|---|
| Entire in vivo test set | All images (n=4814) | 0.156 [0.092, 0.189] | 0.139 [0.068, 0.180] |
| Anatomical regions | Biceps (n=725) | 0.214 [0.118, 0.296] | 0.202 [0.108, 0.287] |
| | Breast (n=435) | 0.179 [0.118, 0.224] | 0.166 [0.106, 0.213] |
| | Calf (n=696) | 0.152 [0.087, 0.175] | 0.133 [0.058, 0.168] |
| | Carotid (n=754) | 0.128 [0.089, 0.159] | 0.115 [0.075, 0.148] |
| | Colon (n=870) | 0.152 [0.094, 0.193] | 0.136 [0.068, 0.183] |
| | Elbow (n=377) | 0.129 [0.087, 0.157] | 0.094 [0.037, 0.133] |
| | Neck (n=203) | 0.128 [0.083, 0.156] | 0.109 [0.053, 0.145] |
| | Thyroid (n=754) | 0.143 [0.079, 0.165] | 0.128 [0.059, 0.155] |
| Participants | 01 (n=667) | 0.148 [0.094, 0.174] | 0.132 [0.068, 0.163] |
| | 02 (n=638) | 0.084 [0.065, 0.097] | 0.050 [0.028, 0.069] |
| | 03 (n=1015) | 0.101 [0.074, 0.123] | 0.083 [0.045, 0.113] |
| | 04 (n=899) | 0.254 [0.174, 0.318] | 0.245 [0.166, 0.310] |
| | 05 (n=986) | 0.183 [0.130, 0.218] | 0.172 [0.120, 0.208] |
| | 06 (n=609) | 0.141 [0.102, 0.164] | 0.126 [0.088, 0.153] |
| Fitzpatrick scale | 2 (n=1566) | 0.209 [0.131, 0.280] | 0.197 [0.120, 0.270] |
| | 3 (n=2610) | 0.141 [0.094, 0.168] | 0.127 [0.076, 0.159] |
| | 4 (n=986) | 0.084 [0.065, 0.097] | 0.050 [0.028, 0.069] |
| Body type | Endomorph (n=1914) | 0.173 [0.094, 0.229] | 0.159 [0.075, 0.217] |
| | Mesomorph (n=1914) | 0.125 [0.080, 0.149] | 0.103 [0.045, 0.137] |
| | Ectomorph (n=986) | 0.183 [0.130, 0.218] | 0.172 [0.120, 0.208] |
| Wavelengths (nm) | 700 (n=166) | 0.142 [0.082, 0.181] | 0.108 [0.034, 0.158] |
| | 710 (n=166) | 0.140 [0.080, 0.178] | 0.111 [0.036, 0.163] |
| | 720 (n=166) | 0.142 [0.079, 0.179] | 0.114 [0.037, 0.164] |
| | 730 (n=166) | 0.142 [0.076, 0.178] | 0.116 [0.035, 0.165] |
| | 740 (n=166) | 0.142 [0.076, 0.183] | 0.118 [0.036, 0.172] |
| | 750 (n=166) | 0.142 [0.077, 0.189] | 0.120 [0.038, 0.180] |
| | 760 (n=166) | 0.144 [0.076, 0.194] | 0.122 [0.041, 0.181] |
| | 770 (n=166) | 0.150 [0.077, 0.200] | 0.130 [0.042, 0.188] |
| | 780 (n=166) | 0.159 [0.080, 0.214] | 0.139 [0.044, 0.203] |
| | 790 (n=166) | 0.167 [0.082, 0.225] | 0.147 [0.047, 0.215] |
| | 800 (n=166) | 0.172 [0.085, 0.238] | 0.153 [0.051, 0.229] |
| | 810 (n=166) | 0.175 [0.088, 0.238] | 0.157 [0.055, 0.227] |
| | 820 (n=166) | 0.178 [0.090, 0.247] | 0.161 [0.058, 0.233] |
| | 830 (n=166) | 0.181 [0.089, 0.245] | 0.165 [0.065, 0.236] |
| | 840 (n=166) | 0.180 [0.092, 0.248] | 0.164 [0.068, 0.241] |
| | 850 (n=166) | 0.185 [0.096, 0.255] | 0.170 [0.076, 0.247] |
| | 860 (n=166) | 0.188 [0.104, 0.260] | 0.173 [0.082, 0.252] |
| | 870 (n=166) | 0.186 [0.109, 0.258] | 0.172 [0.092, 0.250] |
| | 880 (n=166) | 0.181 [0.112, 0.251] | 0.168 [0.097, 0.243] |
| | 890 (n=166) | 0.191 [0.120, 0.263] | 0.178 [0.102, 0.254] |
| | 900 (n=166) | 0.181 [0.122, 0.241] | 0.169 [0.109, 0.234] |
| | 910 (n=166) | 0.148 [0.121, 0.181] | 0.138 [0.111, 0.174] |
| | 920 (n=166) | 0.126 [0.111, 0.142] | 0.117 [0.103, 0.136] |
| | 930 (n=166) | 0.115 [0.103, 0.129] | 0.107 [0.095, 0.121] |
| | 940 (n=166) | 0.130 [0.117, 0.148] | 0.122 [0.107, 0.142] |
| | 950 (n=166) | 0.152 [0.129, 0.178] | 0.146 [0.120, 0.173] |
| | 960 (n=166) | 0.133 [0.105, 0.156] | 0.127 [0.099, 0.153] |
| | 970 (n=166) | 0.123 [0.095, 0.144] | 0.118 [0.089, 0.140] |
| | 980 (n=166) | 0.119 [0.094, 0.138] | 0.114 [0.088, 0.134] |
| Speed of sound (m/s) | 1475 (n=58) | 0.241 [0.188, 0.300] | 0.233 [0.183, 0.291] |
| | 1480 (n=203) | 0.199 [0.127, 0.262] | 0.187 [0.119, 0.252] |
| | 1485 (n=261) | 0.241 [0.179, 0.287] | 0.233 [0.174, 0.276] |
| | 1490 (n=406) | 0.190 [0.121, 0.248] | 0.176 [0.106, 0.235] |
| | 1495 (n=464) | 0.146 [0.077, 0.178] | 0.125 [0.043, 0.166] |
| | 1500 (n=1131) | 0.156 [0.086, 0.193] | 0.137 [0.056, 0.182] |
| | 1505 (n=754) | 0.131 [0.089, 0.160] | 0.112 [0.059, 0.148] |
| | 1510 (n=725) | 0.123 [0.083, 0.148] | 0.105 [0.056, 0.138] |
| | 1515 (n=493) | 0.162 [0.094, 0.192] | 0.149 [0.078, 0.191] |
| | 1520 (n=174) | 0.135 [0.099, 0.161] | 0.124 [0.087, 0.155] |
| | 1525 (n=145) | 0.139 [0.088, 0.192] | 0.129 [0.080, 0.185] |

Extended Data Table 1. Quantitative evaluation of the image quality for deep model-based (DeepMB) and model-based (MB) reconstructions, with different aggregations of the in vivo test dataset. The table shows the mean values and in square brackets the 25$^{th}$ and 75$^{th}$ percentiles of the data residual norms for in-focus images (4814 in vivo sinograms from the test dataset reconstructed with each one optimal SoS values). The number of images in each category (n) is indicated between round brackets.